\documentclass{article}

\usepackage{arxiv}
\usepackage[utf8]{inputenc} % allow utf-8 input
\usepackage[T1]{fontenc}    % use 8-bit T1 fonts
\usepackage{hyperref}       % hyperlinks
\usepackage{url}            % simple URL typesetting
\usepackage{booktabs}       % professional-quality tables
\usepackage{amsfonts}       % blackboard math symbols
\usepackage{nicefrac}       % compact symbols for 1/2, etc.
\usepackage{microtype}      % microtypography
\usepackage{lipsum}		% Can be removed after putting your text content
\usepackage{graphicx}
\usepackage{natbib}
\usepackage{doi}
\usepackage[normalem]{ulem}
\usepackage{amsmath}               % assumes amsmath package installed
  \allowdisplaybreaks[1]           % allow eqnarrays to break across pages
\usepackage{amssymb}               % assumes amsmath package installed 
\usepackage{url}                   % format hyperlinks correctly
\usepackage{rotating}              % allow portrait figures and tables
\usepackage{multirow}              % allows merging of rows in tables
\usepackage{lscape}                % allows pages to be typeset in landscape mode
\usepackage{tabularx}              % allows fixed width tables
\usepackage{verbatim}              % enhanced version of built-in verbatim environment
\usepackage{footnote}              % allows more control over footnote environments
\usepackage{float}                 % allows H option on floats to force here placement
\usepackage{booktabs}              % improve table line spacing
\usepackage[base]{babel}           % required for lisum package
\usepackage{subcaption}            % for multiple sub-figures in a single float
\usepackage{siunitx}               % add SI units
\usepackage{indentfirst} %首行缩进
\usepackage{ragged2e} %两端对齐
\usepackage{caption} %图片标题
\usepackage{nameref}
\usepackage{xcolor}

\title{Thermal Stability and Carrier Recombination Kinetics in InGaN/GaN Multiple Quantum Wells under High-Temperature Annealing}

\date{} 					% Or removing it

\author{Quan-Shan Liu 
\thanks{The following article has been submitted to the Journal of Applied Physics.}
 \\
	Department of Electrical and Electronic Engineering\\
	Photon Science Institute, University of Manchester\\
	Oxford Road, Manchester, M13 9PL, UK \\
	\texttt{quan-shan.liu@manchester.ac.uk} \\
	\And
	Sadia Sheraz\\
	Department of Chemistry\\
	Photon Science Institute, University of Manchester\\
	Oxford Road, Manchester, M13 9PL, UK \\
	\texttt{sadia.sheraz@manchester.ac.uk} \\
	\And
	Sabina Gurung\\
	Department of Physics and Astronomy \\
    Photon Science Institute, University of Manchester \\
	Oxford Road, Manchester, M13 9PL, UK \\
	\texttt{sabina.gurung@manchester.ac.uk} \\
	\And
	Maddison Coke\\
	Department of Electrical and Electronic Engineering \\
    Photon Science Institute, University of Manchester \\
	Oxford Road, Manchester, M13 9PL, UK \\
	\texttt{maddison.coke@manchester.ac.uk} \\
	\And
	Nicholas Lockyer\\
	Department of Chemistry \\
    Photon Science Institute, University of Manchester \\
	Oxford Road, Manchester, M13 9PL, UK \\
	\texttt{nick.lockyer@manchester.ac.uk} \\
	\And
	Richard J. Curry \thanks{Author to whom any correspondence should be addressed.} \\
	Department of Electrical and Electronic Engineering \\
    Photon Science Institute, University of Manchester\\
	Oxford Road, Manchester, M13 9PL, UK \\
	\texttt{richard.curry@manchester.ac.uk} \\
}

\renewcommand{\shorttitle}{Thermal Stability and Carrier Recombination Kinetics in InGaN/GaN Multiple Quantum Wells under High-Temperature Annealing}

\hypersetup{
pdftitle={A template for the arxiv style},
pdfsubject={q-bio.NC, q-bio.QM},
pdfauthor={David S.~Hippocampus, Elias D.~Striatum},
pdfkeywords={First keyword, Second keyword, More},
}

\begin{document}

\maketitle

\begin{abstract}
	The thermal stability and carrier recombination kinetics of an as-received InGaN/GaN multiple quantum well (MQW) structure, capped with a protective AlN thin film, are studied following a series of thermal annealings at temperatures between 500 °C and 1100 °C. Under 325 nm focused laser excitation at room temperature, the sample's photoluminescence (PL) spectrum exhibits three emission bands: an ultraviolet peak at 363 nm, a blue peak at 455 nm, and a yellow peak at 565 nm. We find that the MQW's 455 nm emission is preserved after annealing at 1100 °C. Moreover, room-temperature PL excitation (PLE) and time-resolved PL (TRPL) have been investigated to shed light on the sample's energy-transfer mechanisms and emission decay characteristics. Power-dependent PL spectra analysis shows that the carrier recombination mechanism of the MQW's emission has not been affected by thermal treatment. Rate equation modelling and chromaticity coordinate analysis also show limited thermal impact on the calculated equivalent emission lifetime and emissive colour. Time-of-flight secondary ion mass spectrometry (ToF-SIMS) analysis has been performed, further evidencing the preservation of the MQW structure in the annealed sample.
\end{abstract}

% keywords can be removed
% \keywords{First keyword \and Second keyword \and More}

\section{Introduction}
\label{Section1}
As a flagship wide-bandgap semiconductor in the III–V family, gallium nitride (GaN) has garnered research interest due to its high breakdown voltage and excellent thermal conductivity\citep{ding2019review}. Those advantages make GaN a superior candidate for high-frequency and high-temperature applications. At room temperature, GaN is considered to have a direct band gap of approximately 3.4 eV (in the wurtzite structure) or 3.2 eV (in the zinc-blende structure)\citep{kim1997multiphoton,lei1993heteroepitaxy}. The bandgap of GaN can be readily tuned via alloying to obtain InGaN and AlGaN compositions. As such, GaN and its structural derivatives are widely applied in optoelectronics, including light-emitting diodes (LEDs) and laser diodes\citep{nakamura1994candela,nakamura1998roles}. In particular, the bandgap of InGaN is expected to be adjustable between 0.7 eV (pure InN) and 3.4 eV (pure GaN), allowing light emission across the whole visible range\citep{vurgaftman2001band,vurgaftman2003band}. Therefore, great effort has been devoted to the study of InGaN to develop advanced devices such as highly efficient solar cells and full-spectrum LEDs\citep{fan2023monolithically,matioli2011high,bhuiyan2012ingan}.

Traditional methods for growing InGaN include metal-organic chemical vapour deposition (MOCVD) and molecular beam epitaxy (MBE)\citep{capper2017epitaxial,wang2004molecular}. Although high-quality epilayers can be achieved, the thermal stability remains problematic. For InGaN-based devices, post-growth thermal annealing is widely employed both to activate Mg acceptors in the p-GaN capping layer\citep{nakamura1992thermal,neugebauer1996role}, and to form low-resistance ohmic contacts during subsequent chip fabrication\citep{ho1999low}. However, the thermal budget for these steps must be strictly controlled; elevated annealing temperatures can severely degrade the underlying InGaN active region by inducing indium phase separation\citep{lin2002effects}, interdiffusion\citep{chuo2000effects,cheng2004impact}, and structural defects\citep{chuo2000effects,liu2018influence}. To alleviate the impact of high temperature on InGaN alloys, several solutions originally developed for nitrides may be considered, including: (i) annealing at extremely high nitrogen pressure to suppress thermal decomposition\citep{kuball2000high}; (ii) deposition of an AlN capping layer on the surface before annealing to serve as an atom-diffusion barrier\citep{zolper1996sputtered}; (iii) annealing under a reactive ambient such as NH$_3$ to stabilize the volatile dynamic equilibrium\citep{hernandez2017inxga1}.

In this paper, we report experimental studies of an AlN-capped InGaN/GaN MQW structure, with the aim of investigating the high-temperature annealing impact on its thermal stability and carrier dynamics. Steady-state photoluminescence (PL) and Commission Internationale de l’Eclairage (CIE) chromaticity analysis across all processing stages are conducted to trace the macroscopic optical evolution and colour reproducibility. PL excitation (PLE) and wavelength-selective time-resolved PL (TRPL) measurements are employed to map carrier excitation pathways and transient decay lifetimes of both excitonic and deep-level channels. Furthermore, excitation-power-dependent PL coupled with a two-channel rate-equation model is established to decouple the carrier recombination kinetics and saturation behaviours. Finally, time-of-flight secondary ion mass spectrometry (ToF-SIMS) depth profiling is performed to directly probe the interfacial integrity and elemental distribution within the MQW active region following extreme thermal processing.

\section{Characterisation methods}
\label{Section2}
The sample investigated in this work is as-received wurtzite GaN grown on a sapphire substrate. An MQW InGaN structure is buried within the GaN, revealed using ToF-SIMS characterisation (vide infra). Prior to any thermal annealing, an AlN thin film was deposited on the sample surface via radio frequency sputtering at a deposition rate of $\sim$0.9 nm/s. The AlN film thickness measured using atomic force microscopy (AFM) is $\sim$55.67 ±1.67 nm and shows high transmittance across the visible range (see Supplementary Information: section 1). Following the deposition of the AlN cap, a series of rapid thermal anneals (AnnealSys As-One 100 rapid thermal processor) were performed under a nitrogen atmosphere using a controlled heating ramp rate of 1 °C/s and cooling provided via circulated water-cooling. For each anneal the sample was held at a set maximum temperature for 60 seconds with the chosen maximum temperature being increased from 500 °C to 1100 °C in 100 °C increments (summarised in Table \ref{table:stage definition}). The same AlN/GaN sample underwent each anneal; thus the total annealing received over the course of the study is cumulative. Following this, the sample's UV-Vis absorption spectrum was obtained, yielding an optical bandgap value of 3.34 eV ($\sim$371 nm) via the Tauc relation (see Supplementary Information: section 1).

    \begin{table}[!ht]
        \centering
        \caption{Summary of annealing cycles undertaken.}
        \begin{tabular}{ccccc}
        \hline
            Stage & Definition & Comments  \\ \hline
            1 & Uncapped & -  \\ 
            2 & Capped & deposited with $\sim$55.67 nm AlN \\
            3 & Post-annealed at 500 °C & N$_2$ atmosphere for 60 seconds \\
            4 & Post-annealed at 600 °C & N$_2$ atmosphere for 60 seconds \\
            5 & Post-annealed at 700 °C & N$_2$ atmosphere for 60 seconds \\
            6 & Post-annealed at 800 °C & N$_2$ atmosphere for 60 seconds \\
            7 & Post-annealed at 900 °C & N$_2$ atmosphere for 60 seconds \\
            8 & Post-annealed at 1000 °C & N$_2$ atmosphere for 60 seconds \\
            9 & Post-annealed at 1100 °C & N$_2$ atmosphere for 60 seconds \\
            \hline
        \end{tabular}
        \label{table:stage definition}
    \end{table}

Steady-state (micro) PL spectroscopy was performed on the sample at each of the stages shown in Table \ref{table:stage definition}. A Kimmon IK series He-Cd 325 nm laser was used as the excitation source, corresponding to an energy of $\sim$3.81 eV, which is sufficient for GaN and InGaN excitation. The continuous-wave laser was focused onto the sample surface with the use of a 15X near-ultraviolet objective lens. The effective laser spot diameter was estimated to be $\sim$20 \textmu m using a camera to image it. The PL emission was collected using a Thorlabs fibre-coupled spectrometer (CCS200), and the resulting PL spectra were corrected for the spectral response of the measurement system (including fibre coupling and transmission loss) and the AlN capping layer transmittance. The measurement system correction curve was acquired by comparing the measured spectrum of a Bentham IL1 halogen source (350-4000 nm) with the given spectral irradiance of this lamp as a function of wavelength (see Supplementary Information: section 2). Emission characteristics including peak centre, full width at half maximum (FWHM), and integrated intensity of each emission band observed were extracted from the spectra for quantitative analysis. To enable this fitting, the corrected PL intensity, $I^{PL}(\lambda)$, was multiplied by $\lambda^2$ in order to plot the spectra as a function of photon energy (arising from conversion of the $I^{PL}(\lambda)$ dependence to the $I^{PL}(\hbar\omega)$ dependence)\citep{reshchikov2018two,pelant2012luminescence}. The emission bands were then separated and fitted with Gaussian curves, followed by their integration to obtain the emission intensity.

PLE spectra were recorded using a Fluorolog-3 FL3-22 with a double-grating excitation monochromator and a triple-grating iHR320 spectrometer coupled to an R928P photomultiplier tube for detection. A 450 W Xe lamp was used as the excitation source. The same set-up was also used to measure steady-state PL spectra at excitation wavelengths of 266 nm and 400 nm. The measurements were carried out using a front-face acquisition geometry, and the acquired data were corrected using the built-in instrument functions provided. 

The emission decay dynamics were investigated using TRPL measurements based on the time-correlated single-photon counting (TCSPC) technique. The sample was excited using either the second or third harmonic of a pulsed 80 MHz Ti:Sapphire laser, providing excitation wavelengths of 400 nm and 266 nm, respectively. A pulse picker was used to down-pick the repetition rate, extending the pulse arrival interval to accommodate long (\textmu s) lifetimes. The excitation path utilised an optional focusing lens, which provided a macro-PL configuration. Due to the relatively large spot size, the system yielded bulk-dominated PL decay dynamics akin to an unfocused laser measurement. TRPL data were captured with a monochromator and a microchannel plate photomultiplier tube detector, yielding a total instrument response function of approximately 50 ps. The raw TCSPC data were acquired on a PicoHarp 330 system with a digitised channel resolution (bin width) of either 4 picoseconds for short-lived emission or 2.048 nanoseconds for long-lived emission. 

    \begin{equation}     
        I(t) = A\exp\left[-(t/\tau)^{\delta}\right] + B_g.     
        \label{equ:stretched} 
    \end{equation}

The PL decay curves are fitted using a stretched-exponential model (Equation \ref{equ:stretched})\citep{pophristic1998time}, where $\tau$ is the characteristic decay time and $B_g$ is a constant background. $\delta$ ($0 < \delta \leq 1$) is the stretching (dispersion) exponent describing the system's dimensionality and degree of disorder, where $\delta = 1$ signifies an ideal single-energy discrete transition, and a lower $\delta$ quantifies a broader, continuous distribution of relaxation rates arising from spatial or energetic landscape disorder. Because $\tau$ is not itself the mean decay time when $\delta < 1$, the comprehensive carrier dynamics were quantified using the intensity-weighted average lifetime ($\tau_{\text{avg}}$) as defined in Equation \ref{equ:tauavg}, where $\Gamma$ is the Gamma function.

    \begin{equation}     
        \begin{aligned}
            \langle \tau \rangle &= \frac{\int_0^{\infty} t\,I(t)\,dt}{\int_0^{\infty} I(t)\,dt} = \tau\,\frac{\Gamma(2/\delta)}{\Gamma(1/\delta)}, \quad
            \text{where } \Gamma(x) = \int_0^{\infty} u^{x-1} e^{-u} \, du \quad (x > 0).
        \end{aligned}
        \label{equ:tauavg} 
    \end{equation} 

Fitting was performed by minimising the Poisson deviance ($D$) (Equation \ref{equ:MLE fitting})\citep{maus2001experimental}, where $n_i$ and $m_i$ are the observed and model-predicted counts in channel $i$. Minimising $D$ is equivalent to maximising the Poisson log-likelihood. The goodness-of-fit was assessed from the reduced deviance $D/\nu$ ($\nu$: degrees of freedom, calculated as the number of analysed data channels in the fit minus the number of free parameters being optimised), with values close to unity indicating an adequate fit.
    
    \begin{equation}
        D = 2 \sum_{i} \left[ m_i - n_i + n_i \ln\left(\frac{n_i}{m_i}\right) \right].
        \label{equ:MLE fitting}
    \end{equation}

    \begin{table}[!ht]
        \centering
        \caption{325 nm laser excitation powers used for micro-photoluminescence studies.}
        \begin{tabular}{ccc}
        \hline
            Power received (mW) & Nominal average power density ($\text{W/cm}^2$) & Power percentage (\%)\\ \hline
            6.35 & 2021.27 & 100 \\ 
            3.33 & 1059.97 & 52.4 \\ 
            1.48 & 471.10 & 23.3  \\ 
            0.60 & 190.99 & 9.4 \\ 
            0.43 & 136.87 & 6.8  \\ 
            0.24 & 76.39 & 3.8  \\ 
            57.5 $\times$ 10$^{-3}$ & 18.30 & 0.91  \\ 
            5.04 $\times$ 10$^{-3}$ & 1.60 & 0.08  \\ 
            0.55 $\times$ 10$^{-3}$ & 0.18 & 0.009  \\ \hline
        \end{tabular}
        \label{table:focused laser power}
    \end{table}

Following annealing at 500 °C and 1000 °C, the excitation power dependence of the 325 nm excited PL properties was investigated with the use of neutral-density (ND) filters. These enabled excitation powers to be studied ranging from 6.35 mW (no attenuation) to 0.55 \textmu W, tabulated in Table \ref{table:focused laser power}. To quantify the power dependency of the emission peaks, the corrected PL spectra $I^{PL}(\lambda)$ underwent similar data processing to that described above.

The power-dependent PL spectra were subsequently used for rate equation modelling. Parameters in the model were extracted based on a global optimisation approach. To obtain experimental PL intensities for use, the corrected PL spectra $I^{PL}(\lambda)$ acquired in the power-dependent results were multiplied by $\lambda^{3}$ to convert to photon number. Afterwards, emission peaks in the converted PL spectra were separated and individually fitted with a Gaussian curve to enable intensity integration.

The measured PL spectra were transformed into the CIE 1931 chromaticity coordinates. The tristimulus values $(X, Y, Z)$ were calculated by computing the numerical integration of the measured spectral power distribution $I^{PL}(\lambda)$ with the CIE 1931 2° colour matching functions. Finally, the chromaticity coordinates $(x, y)$ were obtained through the standard normalization $x = X/(X+Y+Z)$ and $y = Y/(X+Y+Z)$.

After all the annealing steps and PL characterisation were complete, ToF-SIMS analysis was performed (J105, Ionoptika Ltd) to investigate the spatial composition of the sample. The measurement was performed by raster scanning a focused C$^{+}_{60}$ cluster ion source operating at 40 keV over a 100 $\times$ 100 \textmu m$^2$ region with a 2 \textmu m spatial resolution. The ion dose per layer was maintained at $\sim$1.13E13 ions/cm$^2$ for a trade-off between sputtering rate and depth profiling quality. Data integration was acquired from 25$\times$25 pixels in a 50 $\times$ 50 \textmu m$^2$ central area of the sputtered crater, to ensure a flat bottom profile and minimise edge artefacts.

Unless otherwise specified, in relevant resulting figures, vertical error bars represent the covariance-propagated uncertainty from the fit covariance matrix. Horizontal error bars represent a conservative estimate based on the numerical resolution of the recorded laser power readings.

\section{Results and discussion}
\label{Section3}

\subsection{Steady-state photoluminescence spectrum analysis}
\begin{figure}[htbp]
      \centering
      \includegraphics[width=1\textwidth]{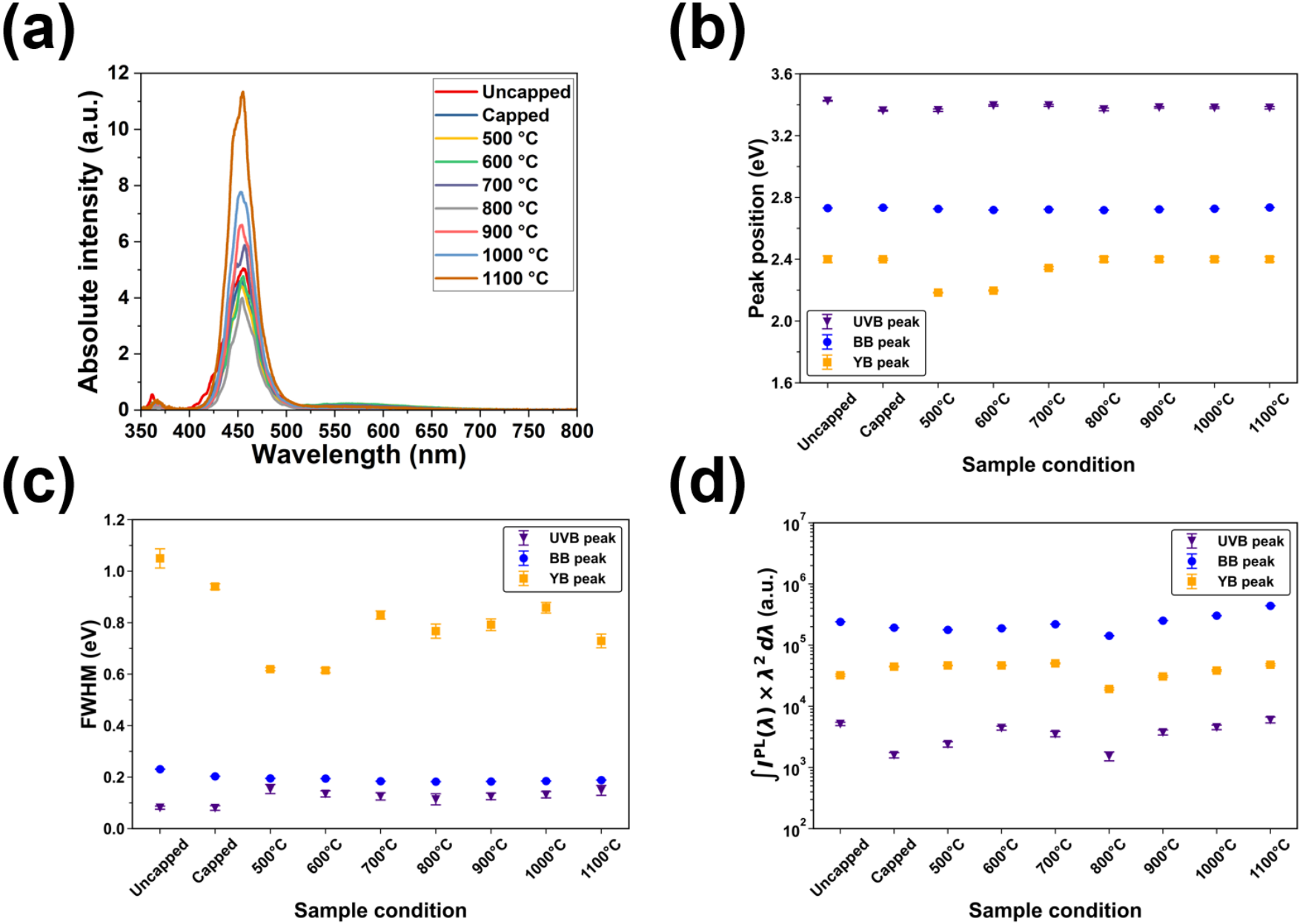}
      \caption{(a) Room temperature PL spectra of the sample at all stages of processing. (b) to (d) The PL peak centre, FWHM and integrated intensity respectively for each emission band obtained at each experimental stage.}
      \label{figure_RT_PL}
    \end{figure}

Figure \ref{figure_RT_PL}(a) shows the room-temperature PL spectra of the sample at all experimental stages obtained using the highest excitation power density. Three emission bands can be observed in the spectra. The ultraviolet band (UVB) of low intensity at $\sim$363 nm is considered to be the near band edge emission of wurtzite GaN\citep{kim1997multiphoton}. The dominant blue band (BB) emission at $\sim$455 nm results from the InGaN/GaN MQW structure. In addition, a broad yellow band (YB) on the low-energy side of the MQW emission can be identified, arising from deep-level defects such as C$_{N}$\citep{reshchikov2014carbon}, C$_{N}$O$_{N}$\citep{demchenko2013yellow}, and V$_{Ga}$ complexes\citep{reshchikov2005luminescence,ogino1980mechanism,neugebauer1996gallium}. Figures \ref{figure_RT_PL}(b) to \ref{figure_RT_PL}(d) show the evolution of key PL parameters (peak centre, FWHM, and integrated intensity) of each band prior to and following AlN deposition, and after each thermal annealing step. 

For the BB PL, the peak centre remains stable across all stages of processing, indicating the excellent thermal stability of the indium composition. The gradual narrowing of the BB PL FWHM suggests improved crystalline quality and a more uniform indium distribution. Meanwhile, the integrated intensity initially decreases before increasing (also plotted using a linear scale in Supporting Information, section 3), which is likely attributed to the sequential activation and subsequent annihilation of non-radiative defects near the MQW structure. The characteristics of the UVB and YB PL show increased variability compared to the BB PL with sample processing. This is possibly due to the relatively low emission intensities which cause less reliable convergence of the fitting curves. In general, no significant peak position shifts were observed for the UVB and YB PL after the final annealing. By contrast, thermal treatment does appear to influence the UVB and YB PL FWHMs and integrated intensities, with their values fluctuating.

Accounting for both quantum confinement effects and the severe internal polarisation fields inherent to strained c-plane nitride interfaces, the $\sim$455 nm BB emission photon energy corresponds to a nominal In molar fraction of approximately 15\%–18\%\citep{alam2017ingan,li2009effect}, varying slightly with the geometric design of the quantum well and barrier layers.

\subsection{Photoluminescence excitation spectrum analysis}
\begin{figure}[htbp]
      \centering
      \includegraphics[width=1\textwidth]{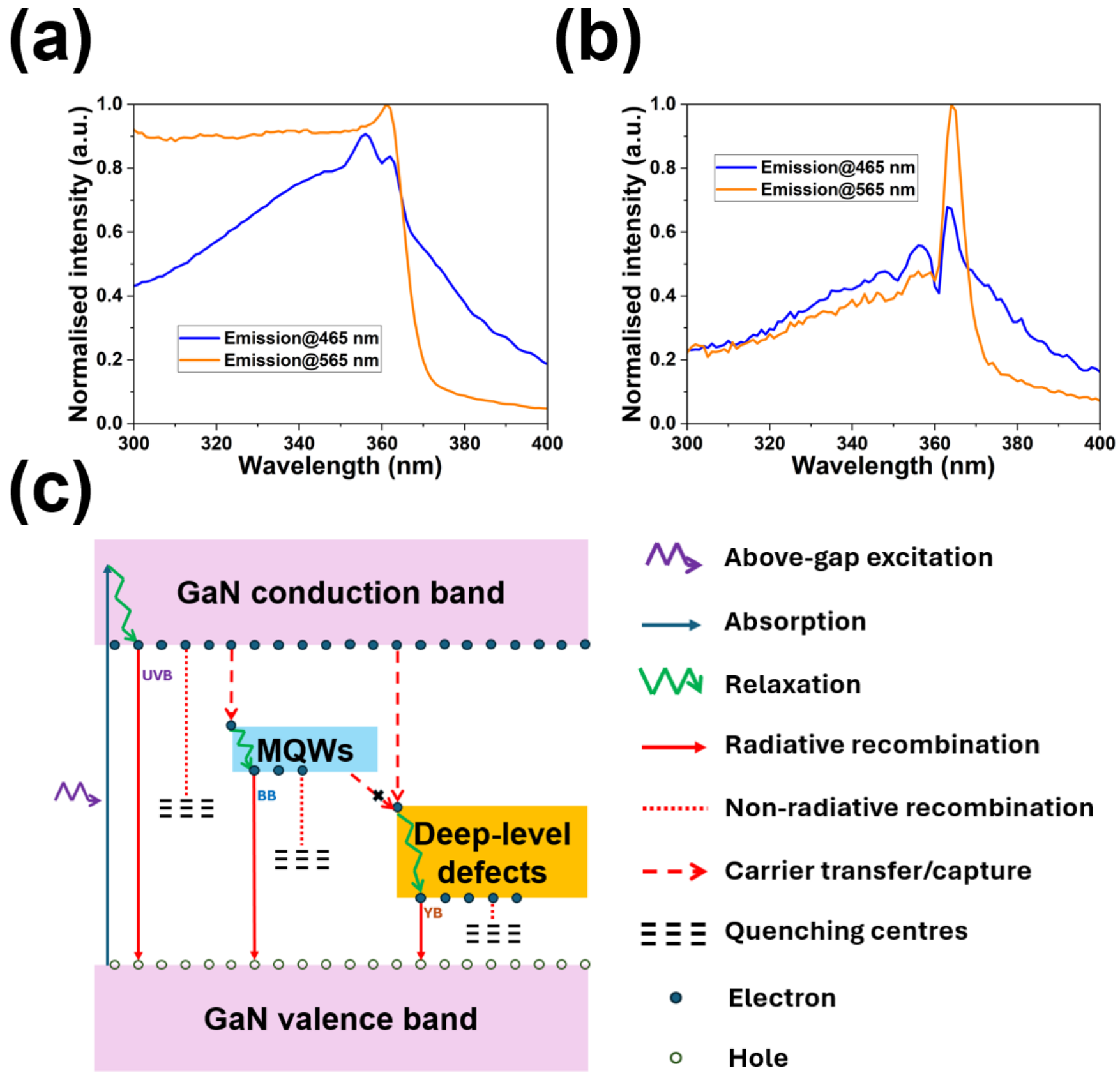}
      \caption{PLE spectra obtained following annealing at (a) 500 °C and (b) 1100 °C. (c) Schematic energy-level diagram for electron excitation and relaxation. }
      \label{figure_PLE}
    \end{figure}

Figures \ref{figure_PLE}(a) and (b) reveal the PLE profiles of the sample post-annealed at 500°C and 1100°C, selectively monitoring the MQW BB ($\lambda_{\text{em}}$ = 465 nm) and the deep-level YB ($\lambda_{\text{em}}$ = 565 nm). The selection of 465 nm rather than 455 nm was due to the BB's redshift under the ultra-low-excitation power density (vide infra) provided by the lamp-based excitation source.

Following annealing at 500 °C (Figure \ref{figure_PLE}(a)), distinct excitation regimes are observed. When monitoring the 565 nm defect emission (orange curve), the PLE spectrum exhibits a notably flat response spanning the deep ultraviolet region (300–360 nm), followed by a steep photoexcitation edge occurring at $\sim$363 nm ($\sim$3.42 eV), which corresponds to the fundamental band gap of bulk GaN. This behaviour implies that free carriers generated anywhere within the GaN can diffuse and be captured by the deep-level defects responsible for the yellow luminescence. The PLE spectrum of the 465 nm MQW emission (blue curve) demonstrates a classic absorption continuum at shorter wavelengths, complemented by a discernible shoulder near the GaN band edge, confirming that the MQW active region relies heavily on 363 nm PL reabsorption, or the transfer of high-energy carriers trapped within the surrounding GaN matrix. At the low excitation powers available for the lamp-based PLE measurements we observed negligible UVB PL at $\sim$363 nm. This indicates that at such ultra-low carrier generation densities, the radiative recombination of free excitons is outcompeted by the non-radiative pathway and the rapid capture kinetics into the MQW and YB-related defect sinks.

A restructuring of the carrier transfer landscape occurs after high-temperature annealing at 1100 °C (Figure \ref{figure_PLE}(b)). It is observed that the above-bandgap flat PLE signal for the YB PL is significantly diminished; instead, a singular, sharp excitation peak is observed at $\sim$363 nm which is mirrored in the BB PLE spectrum. Crucially, as evidenced by the ToF-SIMS depth profiles (discussed below) the InGaN/GaN MQW structure is retained without severe interdiffusion even under 1100°C annealing. Therefore, the appearance of this strong excitonic-like resonance cannot be assigned to structural degradation. Rather, it indicates that the carrier transport and/or carrier capture pathways feeding the YB centres become significantly less efficient. Consequently, the delivery of carriers into both the intact MQWs and the YB-related defect states becomes highly selective, governed by the resonant absorption of free excitons at the GaN band-edge. High-energy, above-bandgap hot carriers generated near the GaN surface now experience enhanced non-radiative recombination before reaching the active region. The excitation of the MQW PL at 465 nm appears to be similar to that observed previously.

It is also noticed that for excitation wavelengths longer than 363 nm, the excitation efficiency of YB emission is consistently lower than that of the BB. This may be explained by the YB PL excitation being dominated by carrier capture mechanisms, which is a less efficient process than the formation of excitons within the MQWs. A schematic energy-level diagram is presented in Figure \ref{figure_PLE}(c) to illustrate the electron excitation and relaxation pathways possible within the sample.

\subsection{Time-resolved emission decay analysis}

\begin{figure}[htbp]
      \centering
      \includegraphics[width=1\textwidth]{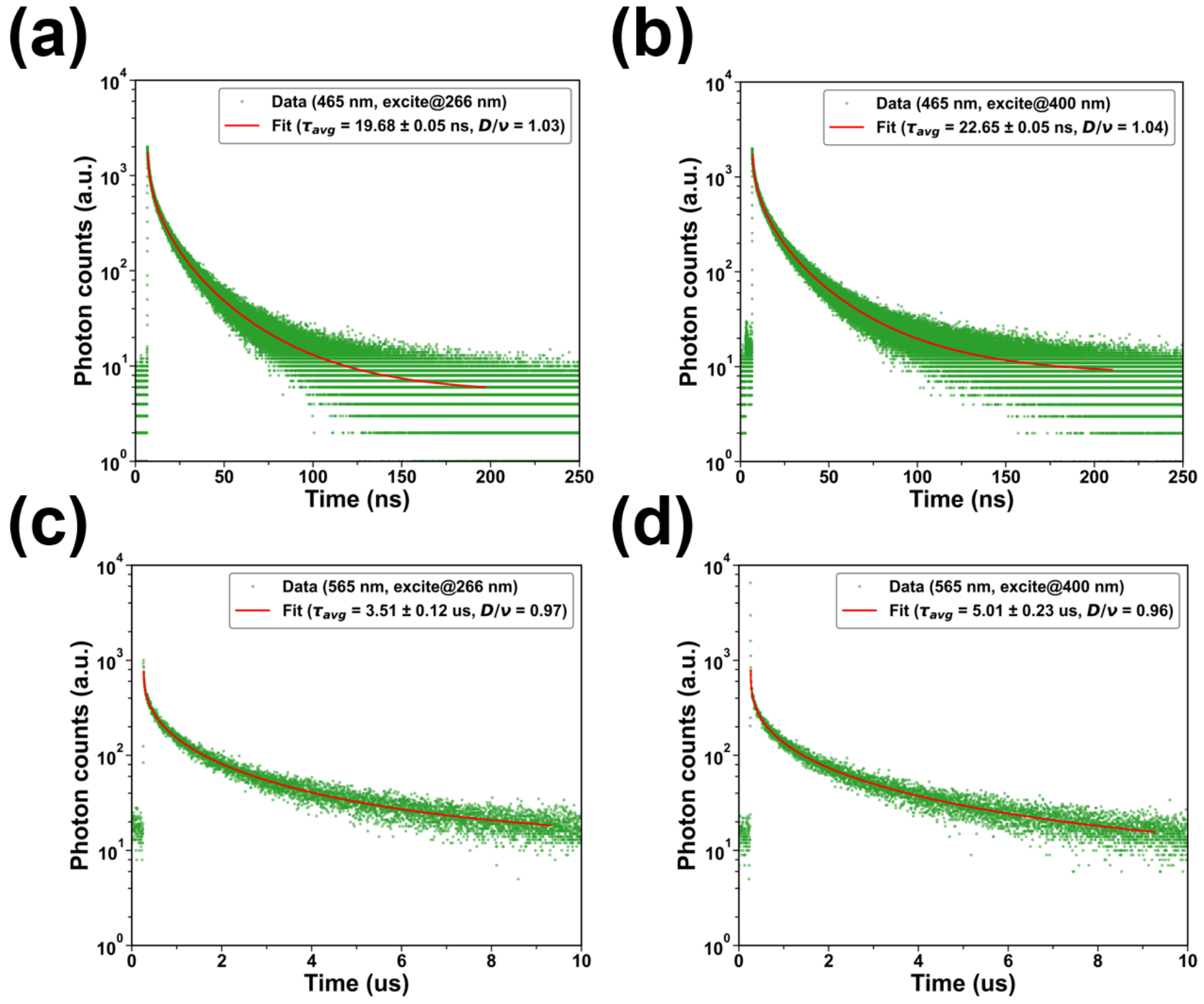}
      \caption{PL decay transient obtained following annealing at 1100 °C for the BB ($\lambda_{\text{em}} = 465\text{ nm}$) PL obtained using an excitation wavelength of (a) 266 nm and (b) 400 nm, and the YB ($\lambda_{\text{em}} = 565\text{ nm}$) PL decay transient obtained at an excitation wavelength of (c) 266 nm and (d) 400 nm.}
      \label{figure_lifetime}
    \end{figure}

To gain deeper insight into the recombination kinetics TRPL measurements were executed utilising the TCSPC technique. Figure \ref{figure_lifetime} illustrates the 1100 °C post-annealed PL decay profiles fitted using a stretched exponential framework. The BB ($\lambda_{\text{em}} = 465\text{ nm}$) and YB ($\lambda_{\text{em}} = 565\text{ nm}$) decay transients are shown under above-bandgap ($\lambda_{\text{ex}} = 266\text{ nm}$) and below bandgap (direct MQW) ($\lambda_{\text{ex}} = 400\text{ nm}$) pulsed excitation. The calculated intensity-weighted average lifetimes ($\tau_{\text{avg}}$) along with their corresponding goodness-of-fit metrics ($D/\nu$) are presented within the legends of subplots.

For the 465 nm BB emission, a small excitation-wavelength dependence is observed. Under direct MQW excitation ($\lambda_{\text{ex}} = 400\text{ nm}$), where photocarriers are generated by direct absorption within the InGaN confined states, the emission exhibits an average lifetime of $\tau_{\text{avg}}$ = 22.65 ±0.05 ns (Figure \ref{figure_lifetime}(b)). Increasing the excitation energy to the above-gap ultraviolet regime ($\lambda_{\text{ex}} = 266\text{ nm}$) causes a noticeable reduction of the average lifetime to $\tau_{\text{avg}}$ = 19.68 ±0.05 ns (Figure \ref{figure_lifetime}(a)). To examine whether this difference originates from changes in the emitting MQW population, steady-state PL spectra were acquired under both excitation wavelengths (see Supporting Information, section 4). 

After normalisation, both spectra exhibit nearly identical MQW emission. This indicates that both excitation schemes probe essentially the same MQW ensemble, although different carrier distributions within the localised states cannot be completely discounted. Intriguingly, the stretching exponent remains almost invariant between the two pumping configurations ($\delta$ = 0.489 ±8.89 $\times 10^{-4}$ for 266 nm versus $\delta$ = 0.495 ±8.88 $\times 10^{-4}$ for 400 nm). This near invariance of $\delta \sim 0.49$ suggests that the intrinsic energetic disorder of the MQWs is independent of the excitation pathway. Consequently, the observed lifetime variation is more likely associated with differences in the initial carrier generation and capture conditions than with any modification of the recombination mechanism itself.

A sharp contrast in PL lifetime is observed for the YB emission (Figures \ref{figure_lifetime}(c) and \ref{figure_lifetime}(d)), where the lifetime is found to be in the microsecond regime, evidencing deep-level trap-assisted transitions. The YB emission transient profiles reveal a greater relative change in both average lifetime and stretching topology for the different excitation wavelengths. Under above-barrier 266 nm pumping, the fitting yields $\tau_{\text{avg}}$ = 3.51 ±0.12\ \textmu s with a stretching exponent of $\delta$ = 0.403 ±5.14 $\times 10^{-3}$. Conversely, direct-MQW 400 nm pumping significantly stretches the decay tail, driving $\delta$ down to 0.360 ±5.07 $\times 10^{-3}$ and increasing the average lifetime to $\tau_{\text{avg}}$ = 5.01 ±0.23 \textmu s. Unlike the MQW emission, the yellow luminescence originates from deep-level centres distributed throughout the GaN matrix rather than from a spatially fixed active region. Consequently, changing the excitation wavelength also changes the spatial carrier-generation profile. Because the optical absorption coefficient of GaN increases rapidly above the bandgap while remaining much lower below the band edge (see Supplementary Information: section 1), the optical penetration depth is expected to differ substantially between 266 nm and 400 nm. 

Under above-gap excitation (266 nm), photocarriers are generated predominantly near the sample surface, where competition from surface-related non-radiative processes is expected to be strongest. By contrast, the substantially larger optical penetration depth at 400 nm allows carriers to populate a deeper distribution of defect centres before recombination. This interpretation is also consistent with the steady-state PL spectra (see Supporting Information, section 4), where the relative YB intensity is significantly enhanced under 266 nm excitation compared with 400 nm excitation, indicating that the excitation pathway modifies the population of defect centres participating in the emission process. The accompanying reduction of $\delta$ therefore most likely reflects changes in the defect distribution arising from the different excitation depths rather than an intrinsic modification of the defect states themselves.

For completeness, Bi-exponential decays were also fitted to the PL decay as used elsewhere \citep{wang2017consistency}; however in all cases the fit quality did not match that of the stretched exponential fitting (Supplementary Information: section 4). We therefore adopt the use of the stretched exponential model in this work.

\subsection{Power-dependent photoluminescence spectrum analysis}

\begin{figure}[htbp]
      \centering
      \includegraphics[width=1\textwidth]{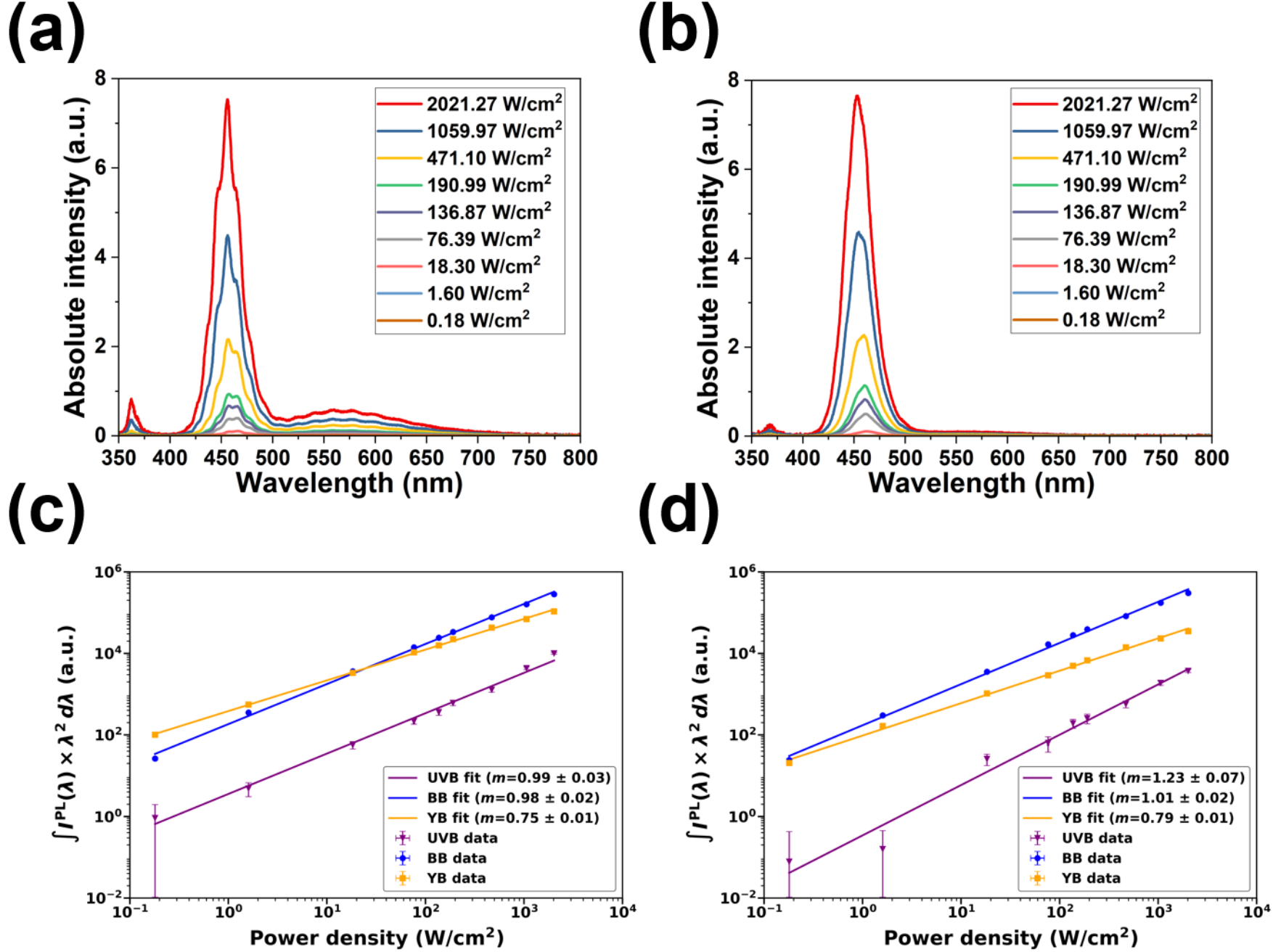}
      \caption{Excitation power density dependence of 325 nm excited PL obtained after (a) 500 °C and (b) 1000 °C annealing.  Log-Log plot of the integrated PL intensity as a function of excitation power density for each emission peak obtained after (c) 500 °C and (d) 1000 °C annealing. Lower error bars of some data were truncated where they extend below the axis limit on the logarithmic scale.}
      \label{figure_PD_PL}
    \end{figure}

Figures \ref{figure_PD_PL}(a) and \ref{figure_PD_PL}(b) present the (micro) PL excitation power density dependence following annealing at 500 °C and 1000 °C respectively. At low excitation power densities, a significant portion of photo-generated carriers are captured by structural defects, leading to non-radiative recombination dominating. Consequently, the steady-state carrier density within the InGaN/GaN MQW is severely reduced. This depletion of carriers weakens the screening effect against the internal piezoelectric field\citep{takeuchi1997quantum,chichibu2002localized}, causing the energy bands to bend. As a result, the peak position of BB emission undergoes a red shift from \textasciitilde 455 nm to \textasciitilde 462 nm due to the enhanced quantum-confined Stark effect (QCSE)\citep{bernardini1997spontaneous}. This mechanism has also been evidenced in strained InGaN quantum wells under low-injection conditions where the polarisation field remains unshielded\citep{peng1999piezoelectric}. Radiative recombination thus preferentially occurs from lower energy states, including the YB and the long-wavelength tail of the BB.

The integrated UVB, BB and YB PL intensity ($\int I^{\mathrm{PL}}(\lambda) \times \lambda^2 \, d\lambda$) as a function of 325 nm excitation power density ($P$) was fitted using the power-law relation $\int I^{\mathrm{PL}}(\lambda) \times \lambda^2 \, d\lambda \propto P^{m}$, as displayed in Figures \ref{figure_PD_PL}(c) and \ref{figure_PD_PL}(d). Following annealing at 500 °C both the UVB and BB display a linear power dependence ($m$ $\approx$ 1.0) suggesting a regime dominated by exciton emission, including free and bound excitons with pre-saturated Shockley-Read-Hall (SRH) non-radiative recombination centres\citep{schmidt1992excitation}. Meanwhile, the YB exhibits a sublinear slope of 0.75. This behaviour is characteristic of free-to-bound or donor-acceptor pair (DAP) recombination, which is governed by the carrier saturation of finite deep-level states\citep{schmidt1992excitation,reshchikov2005luminescence}. 

Subsequent multiple annealing treatments up to 1000 °C alter the recombination landscape. The UVB slope is increased to 1.23, serving as clear evidence of the proliferation of unsaturated SRH non-radiative recombination centres induced by GaN thermal decomposition at 1000 °C\citep{king1998cleaning}. In contrast, the BB PL excitation power density dependence preserves a near-linear slope ($m$ $\approx$ 1.01). This thermal resilience stands in stark contrast to previous literature where uncapped InGaN structures suffered severe catastrophic degradation via massive indium phase separation and interdiffusion \citep{lin2002effects, chuo2000effects}. In the sample studied here, this stability is prolonged by two synergistic factors. The deposited thin AlN capping layer provides a robust thermodynamic barrier that efficiently suppresses the thermal decomposition of the sub-surface matrix and alleviates nitrogen/indium out-diffusion. Kinetically, the brief temporal window inherent to the rapid thermal annealing (RTA) configuration strictly constrains the long-range spatial interdiffusion of In/Ga atoms across the heterostructure interfaces. Consequently, the micro-scale compositional fluctuations are preserved, wherein the strong carrier localisation effect within indium-rich clusters continuously acts as an effective shield, preventing carriers from being captured by the surrounding defect network. 

The thermal evolution of the YB provides insight into the defect dynamics induced by annealing. Upon elevating the annealing temperature to 1000 °C, the integrated PL intensity of the YB is decreased (see Figures \ref{figure_PD_PL}(a) and \ref{figure_PD_PL}(b), and Supporting Information, section 3). This may result from either the reduction of active emission centres via annealing out or partial passivation, or a proliferation of thermally propagated non-radiative defect networks. Any of these factors may lead to a reduced radiative recombination probability, accompanied by a subtle upward shift in the YB power-law exponent from 0.75 to 0.79. Throughout this process, the primary MQW emission remains consistently dominant with a near-linear power dependence. Consequently, the carrier flux trapped into the remaining YB states is heavily suppressed, alleviating phase-space filling of the YB channel within the measured power density window and pushing its kinetic response back toward the linear regime.

To further elucidate the carrier localisation mechanism, power density-dependent PL measurements were additionally conducted under an ultra-low-excitation regime by removing the focusing objective lens (see Supplementary Information: section 5). Under this significantly reduced excitation power density, any UVB emission was either completely quenched by non-radiative pathways or fell below the detection limit. The slope of the BB PL intensity as a function of excitation power density experiences a notable reduction from 1.11 to 1.02 between annealing at 500 °C  and 1000 °C respectively. This linearising trend under ultra-low intensity excitation suggests that the MQWs remain thermodynamically robust at 1000 °C, rather than collapsing into detrimental macroscopic phase separation upon high-temperature annealing. For example, the formation of nanoscopic indium-rich compositional fluctuations (or cluster-induced localisation centres) would produce deep localised states that would efficiently capture and confine carriers. Meanwhile, the slope of the YB PL intensity dependence on excitation power density increases from 0.75 to 0.84, reflecting that these defect states are not saturated under such low power density excitation.

\subsection{Rate equation modelling analysis}
To quantitatively evaluate the competitive carrier dynamics and recombination coefficients within the multi-energy level system, a self-consistent rate equation model was constructed to simulate the excitation power density dependence of the collective PL emission from the three bands. As captured in the steady-state PL profiles, the total emission spectrum is fundamentally comprised of three components: the UVB, BB, and YB PL. The power-law scaling analysis ($\int I^{\mathrm{PL}}(\lambda) \times \lambda^2 \, d\lambda \propto P^{m}$) yields a scaling exponent of $m \gtrsim 1$ for both the UVB and BB PL, verifying that both emissions are primarily governed by radiative excitonic recombination. Given that the intensity of the UVB PL is orders of magnitude weaker than that of the dominant BB PL while exhibiting similar kinetic scaling, we combine the UVB and BB emissions into a single unified excitonic radiative channel to reduce the number of redundant fitting parameters and mitigate model over-parameterisation. Consequently, the conceptual landscape under laser excitation is formulated as follows: photons pump free carriers into an effective upper excitonic state (representing the combined (UVB+BB) reservoir). From this excited state, carriers undergo either direct radiative recombination (yielding the composite band-edge and MQW emission), non-radiative SRH recombination, or are irreversibly captured by a finite density of deep-level defect states that subsequently feed the saturable YB radiative transitions.

To better describe the carrier dynamics, a dual-channel model has been created consisting of Equations \ref{equ:BB} and \ref{equ:YB}. The binary nonlinear equation set is described by two coupled rate equations under steady-state conditions ($\mathrm{d}N_{1(2)}/\mathrm{d}t = 0$), where $N_{1(2)}$ denotes the population of excited carriers in the UVB+BB (YB) state, and $t$ is time. Other parameters are the generation rate of photo-induced carriers ($G$), radiative/non-radiative recombination rates of the (UVB+BB) state ($k_{r1/nr1}$), radiative/non-radiative recombination rates of the YB state ($k_{r2/nr2}$), carrier capture rate from (UVB+BB) to YB ($k_{capture}$), and maximum allowable population of excited carriers at defect states ($N_{defect}$). Among them, $G$ is assumed to change proportionally with the laser power density $P$ and is determined by an equivalent scaling constant $\beta$ ($G = P\beta$). Channel-specific lifetimes and radiative efficiencies are respectively defined in Equations \ref{equ:emission lifetime} and \ref{equ:emission efficiency}. Here, $\tau$ and $\eta$ represent the intrinsic lifetime and radiative efficiency specifically within the designated recombination channel, rather than the overall effective lifetime and quantum efficiency of the entire system.

    \begin{equation}
    dN_1/dt = G - \left( k_{r1} + k_{nr1} + k_{capture} \left( N_{defect} - N_2 \right) \right) N_1 = 0,
    \label{equ:BB}
    \end{equation}

    \begin{equation}
    dN_2/dt = k_{capture} N_1 \left( N_{defect} - N_2 \right) - \left( k_{r2} + k_{nr2} \right) N_2 = 0.
    \label{equ:YB}
    \end{equation}

    \begin{align}
    \tau_1 &= \frac{1}{k_{r1}+k_{nr1}}, \qquad
    \tau_2 = \frac{1}{k_{r2}+k_{nr2}}. \label{equ:emission lifetime}
    \end{align}

    \begin{align}
    \eta_{UVB+BB} &= \frac{k_{r1}}{k_{r1}+k_{nr1}}, \qquad
    \eta_{YB} = \frac{k_{r2}}{k_{r2}+k_{nr2}}. \label{equ:emission efficiency}
    \end{align}

Subsequently, the modelled PL integrated intensities are assumed to scale linearly with the product of the radiative recombination rate and population of excited carriers (Equation \ref{equ:emission intensity}). Prefactor constants related to photon energy, collection efficiency, and detector response are absorbed into the scaling constant $\beta$ and are therefore omitted here. 

    \begin{align}
    \int I\mathrm{_{UVB+BB}^{Model}} \times \lambda^3 \, d\lambda &= k_{r1} N_1 , \qquad
    \int I\mathrm{_{YB}^{Model}} \times \lambda^3 \, d\lambda = k_{r2} N_2. \label{equ:emission intensity}
    \end{align}

    \begin{equation}
    \mathcal{L}_{\mathrm{total}}
    =
    \mathcal{L}_{\mathrm{data}}
    +
    \alpha \, \mathcal{L}_{\mathrm{penalty}}. \label{equ:total loss}
    \end{equation}

Numerical calculation can then be performed by fitting the modelled PL integrated intensities ($\int I\mathrm{_{UVB+BB}^{Model}} \times \lambda^3 \, d\lambda$ and $\int I\mathrm{_{YB}^{Model}}\times \lambda^3 \, d\lambda$) to the experimental values ($\int I\mathrm{_{UVB+BB}^{Exp}}\times \lambda^3 \, d\lambda$ and $\int I\mathrm{_{YB}^{Exp}}\times \lambda^3 \, d\lambda$). This was done by employing a differential evolution algorithm to minimise the total loss function given in Equation \ref{equ:total loss}. The total loss function $\mathcal{L}_{\mathrm{total}}$ combines a data-fitting term $\mathcal{L}_{\mathrm{data}}$ and a penalty term $\mathcal{L}_{\mathrm{penalty}}$, weighted by a coefficient $\alpha$. $\mathcal{L}_{\mathrm{data}}$ is the weighted sum of each emission channel's squared normalised residuals between the experimental and modelled PL integrated intensities across all laser powers, as defined in Equations \ref{equ:residual define} to \ref{equ:data loss}. Normalisation is used to balance any significant difference in order of magnitude between the $\int I\mathrm{_{UVB+BB}^{Exp}}\times \lambda^3 \, d\lambda$ and $\int I\mathrm{_{YB}^{Exp}}\times \lambda^3 \, d\lambda$. The weight function $w_i$ can additionally be used to control the relative contribution of data at different excitation power densities; in this work, uniform weighting ($w_i=1$) is adopted. Because each channel is normalised by a single constant, rather than by the point-wise experimental uncertainty, this means that data points at low power density where $\int I^{\mathrm{Exp}}\times \lambda^3 \, d\lambda$ is intrinsically small contribute proportionally less to $\mathcal{L}_{\mathrm{data}}$ than data points at high power density.

    \begin{align}
        r_{\mathrm{UVB+BB}}(P_i) &=
        \frac{
        \int I^{\mathrm{Model}}_{\mathrm{UVB+BB}}\times \lambda^3 \, d\lambda(P_i)
        -
        \int I^{\mathrm{Exp}}_{\mathrm{UVB+BB}}\times \lambda^3 \, d\lambda(P_i)
        }{
        \max\!\left(\int I^{\mathrm{Exp}}_{\mathrm{UVB+BB}}\times \lambda^3 \, d\lambda \right)
        }, \label{equ:residual define} \\
        r_{\mathrm{YB}}(P_i) &=
        \frac{
        \int I^{\mathrm{Model}}_{\mathrm{YB}}\times \lambda^3 \, d\lambda(P_i)
        -
        \int I^{\mathrm{Exp}}_{\mathrm{YB}}\times \lambda^3 \, d\lambda(P_i)
        }{
        \max\!\left(\int I^{\mathrm{Exp}}_{\mathrm{YB}}\times \lambda^3 \, d\lambda \right)
        },\\
        \mathcal{L}_{\mathrm{data}} &=
        \sum_{i=1}^{n=9}
        w_i^2
        \left[
        r_{\mathrm{UVB+BB}}^2(P_i)
        +
        r_{\mathrm{YB}}^2(P_i)
        \right]. \label{equ:data loss}
    \end{align}

Except for the data fitting loss, an additional penalty term $\mathcal{L}_{\mathrm{penalty}}$ is defined in Equations \ref{equ:penalty define} and \ref{equ:penalty loss}. An inner soft window $[\tau_i^{s-},\tau_i^{s+}]$ and a wider outer hard window $[\tau_i^{h-},\tau_i^{h+}]$ have been established to ensure that all equivalent lifetimes fall into physically reasonable ranges (vide infra). The coefficient $\alpha$ allows tunable emphasis on the penalty term and is set to 0.05 in this work. A group of parameters in Equations \ref{equ:BB} and \ref{equ:YB} ($k_{r1}, k_{nr1}, k_{capture}, k_{r2}, k_{nr2}, N_{defect}$, and $\beta$) can be extracted once the total loss function is minimised. Afterwards, the $N_{1(2)}$ can be calculated by numerically solving the binary nonlinear equation set at different power values. 

    \begin{equation}s
        P(\tau_i) =
        \begin{cases}
        0, & \tau_i^{s-} \le \tau_i \le \tau_i^{s+} \\[4pt]
        \left(\dfrac{\tau_i^{s-}-\tau_i}{\tau_i^{s-}}\right)^{2}, & \tau_i^{h-} \le \tau_i < \tau_i^{s-} \\[8pt]
        \left(\dfrac{\tau_i-\tau_i^{s+}}{\tau_i^{s+}}\right)^{2}, & \tau_i^{s+} < \tau_i \le \tau_i^{h+} \\[8pt]
        \left(\dfrac{\tau_i^{s-}-\tau_i^{h-}}{\tau_i^{s-}}\right)^{2} + w_h\left(\dfrac{\tau_i^{h-}-\tau_i}{\tau_i^{h-}}\right)^{2}, & \tau_i < \tau_i^{h-} \\[8pt]
        \left(\dfrac{\tau_i^{h+}-\tau_i^{s+}}{\tau_i^{s+}}\right)^{2} + w_h\left(\dfrac{\tau_i-\tau_i^{h+}}{\tau_i^{h+}}\right)^{2}, & \tau_i > \tau_i^{h+}
        \end{cases}
        \label{equ:penalty define}
    \end{equation}
    
    \begin{equation}
        \mathcal{L}_{\mathrm{penalty}} = P(\tau_1) + P(\tau_2).
        \label{equ:penalty loss}
    \end{equation}

To assess the algorithm's robustness, the global optimisation was repeated 5 times with randomised initial conditions, providing a statistical insight (mean value and standard deviation) of the extracted parameters. A group of parameters which collectively cause the minimal total loss among 5 simulation runs is obtained as the best-fit parameters. The resulting $N_{1(2)}$ values are used to finalise $\int I\mathrm{_{(UVB+BB)/YB}^{Model}}\times \lambda^3 \, d\lambda$. Finally, the lifetimes and radiative efficiencies of the two emission channels can be evaluated using the model output. It should be noted that a fundamental distinction exists between the steady-state lifetimes modelled under continuous-wave excitation and the transient differential lifetimes captured via pulsed techniques, owing to inherent discrepancies in phase-space filling and defect saturation behaviours. Nevertheless, our experimental TCSPC PL lifetimes provide an indispensable physical foundation, proving unequivocally that the MQW excitonic recombination ($\tau_1$) operates on a nanosecond timescale, whereas the deep-level defect transition ($\tau_2$) extends into the microsecond regime. 

Herein, the parameter search space for the differential-evolution optimiser was intentionally set from 0.5 to 500 ns for $\tau_1$ and from 0.5 to 500 \textmu s for $\tau_2$, respectively. Within this broad search space, the soft window $[10\, ns,100\, ns]/[5\,\mu s,100\,\mu s]$, consistent with the order of magnitude of the TCSPC decay lifetimes, is treated as fully plausible (zero penalty). Beyond this window but within a wider outer hard window $[1\, ns,200\, ns]/[1\,\mu s,200\,\mu s]$, solutions are permitted but increasingly disfavoured. Beyond the outer window, solutions are treated as physically implausible and penalised much more steeply. The two constructs serve distinct roles: the parameter bounds define where the optimiser is permitted to search (an algorithmic requirement of global optimisation, not itself a physical judgement), while the penalty encodes a graded physical belief about which region within that space is more plausible (a distinction a single uniform bound cannot express).

\begin{table}[htbp]
    \centering
    \caption{Overview of the main parameters defined in the rate equation model.}
    \begin{tabularx}{\textwidth}{c X c c}
    \hline
    Symbol & Description & Unit & Initial setting \\
    \hline
    $N_1$ & Excited carrier population in the UVB+BB channel & dimensionless & -- \\
    $N_2$ & Excited carrier population in the YB channel & dimensionless & -- \\
    $N_{defect}$ & Maximum allowable population of excited carriers at defect states & dimensionless & (5E2, 1E4) \\
    $k_{r1}$ & Radiative recombination rate (UVB+BB) & s$^{-1}$ & (1E6, 1E9) \\
    $k_{nr1}$ & Non-radiative recombination rate (UVB+BB) & s$^{-1}$ & (1E6, 1E9) \\
    $k_{r2}$ & Radiative recombination rate (YB) & s$^{-1}$ & (1E3, 1E6) \\
    $k_{nr2}$ & Non-radiative recombination rate (YB) & s$^{-1}$ & (1E3, 1E6) \\
    $k_{\mathrm{capture}}$ & Carrier capture rate from (UVB+BB) to YB & s$^{-1}$ & (1E4, 2E5) \\
    $\tau_1$ & Equivalent carrier lifetime (UVB+BB) & s & -- \\
    $\tau_2$ & Equivalent  carrier lifetime (YB) & s & -- \\
    $\eta_{\mathrm{UVB+BB}}$ & Equivalent radiative efficiency (UVB+BB) & dimensionless & -- \\
    $\eta_{\mathrm{YB}}$ & Equivalent radiative efficiency (YB) & dimensionless & -- \\
    $G$ & Generation rate of photo-generated carriers & s$^{-1}$ & -- \\
    $\beta$ & Equivalent scaling constant relating laser power density to $G$ & cm$^{2}$mW$^{-1}$s$^{-1}$ & (1E2, 1E9) \\
    $P$ & Laser power density & mWcm$^{-2}$ & Table \ref{table:focused laser power} \\
    $\int I_{\mathrm{UVB+BB}}^{\mathrm{Model}}\times \lambda^3 \, d\lambda$ & Modelled PL integrated intensity (UVB+BB) & arbitrary units & -- \\
    $\int I_{\mathrm{YB}}^{\mathrm{Model}}\times \lambda^3 \, d\lambda$ & Modelled PL integrated intensity (YB) & arbitrary units & -- \\
    $\int I_{\mathrm{UVB+BB}}^{\mathrm{Exp}}\times \lambda^3 \, d\lambda$ & Experimental PL integrated intensity (UVB+BB) & arbitrary units & -- \\
    $\int I_{\mathrm{YB}}^{\mathrm{Exp}}\times \lambda^3 \, d\lambda$ & Experimental PL integrated intensity (YB) & arbitrary units & -- \\
    $\mathcal{L}_{\mathrm{total}}$ & Total loss function & dimensionless & -- \\
    $\mathcal{L}_{\mathrm{data}}$ & Weighted sum of squared normalised residuals & dimensionless & -- \\
    $r_{\mathrm{UVB+BB}}(P_i)$ & Normalised residual at the $i^{\mathrm{th}}$ power density (UVB+BB) & dimensionless & -- \\
    $r_{\mathrm{YB}}(P_i)$ & Normalised residual at the $i^{\mathrm{th}}$ power density (YB) & dimensionless & -- \\
    $w_i$ & Residual weighting factor & dimensionless & 1 \\
    $\mathcal{L}_{\mathrm{penalty}}$ & Lifetime penalty term & dimensionless & -- \\
    $[\tau_1^{s-},\tau_1^{s+}]$ & Inner soft penalty window of $\tau_1$ & s & [1E-8, 1E-7] \\
    $[\tau_1^{h-},\tau_1^{h+}]$ & Outer hard penalty window of $\tau_1$ & s & [1E-9, 2E-7] \\
    $[\tau_2^{s-},\tau_2^{s+}]$ & Inner soft penalty window of $\tau_2$ & s & [5E-6, 1E-4] \\
    $[\tau_2^{h-},\tau_2^{h+}]$ & Outer hard penalty window of $\tau_2$ & s & [1E-6, 2E-4] \\
    $w_h$ & Weighting factor for hard penalty window & dimensionless & 30 \\
    $\alpha$ & Global penalty weighting factor & dimensionless & 0.05 \\
    \hline
    \label{table:model parameters}
    \end{tabularx}
\end{table}

\begin{figure}[htbp]
      \centering
      \includegraphics[width=1\textwidth]{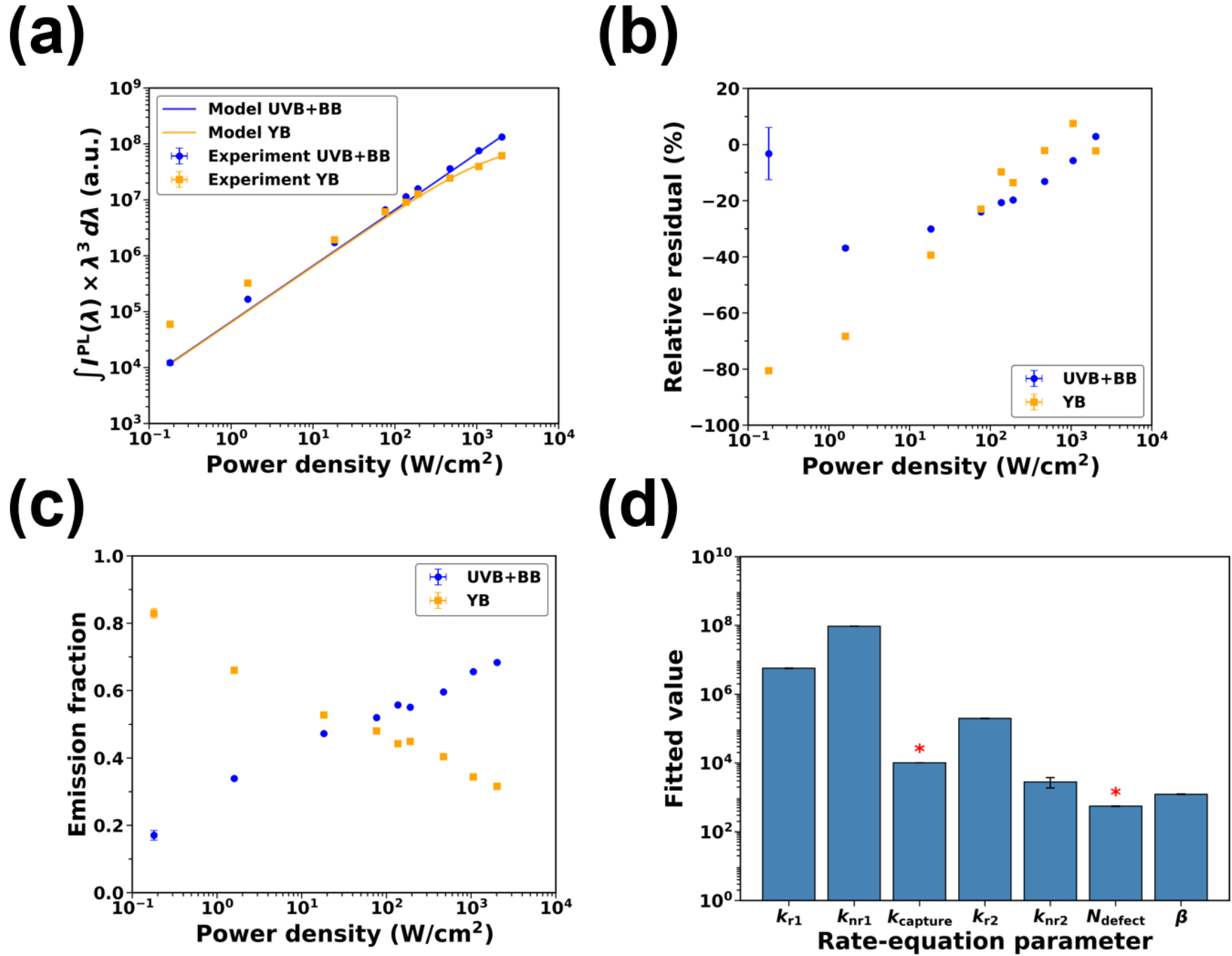}
      \caption{Summary of rate-equation modelling for PL obtained after 500 °C annealing. (a) Comparison between the emission channels' experimental PL integrated intensities (data points) and the best fit obtained using the rate-equation model (lines) as a function of excitation power density. (b) Relative residuals between the emission channels' experimental and modelled PL integrated intensities. (c) Relative channel contribution of the experimental PL integrated intensities. (d) Mean fitting parameter values and standard deviation obtained from repeated differential evolution runs. An `*' identifies a parameter whose best-fit value lies within 5\% of a search-space boundary.}
      \label{figure_REM_500}
    \end{figure}

\begin{figure}[htbp]
      \centering
      \includegraphics[width=1\textwidth]{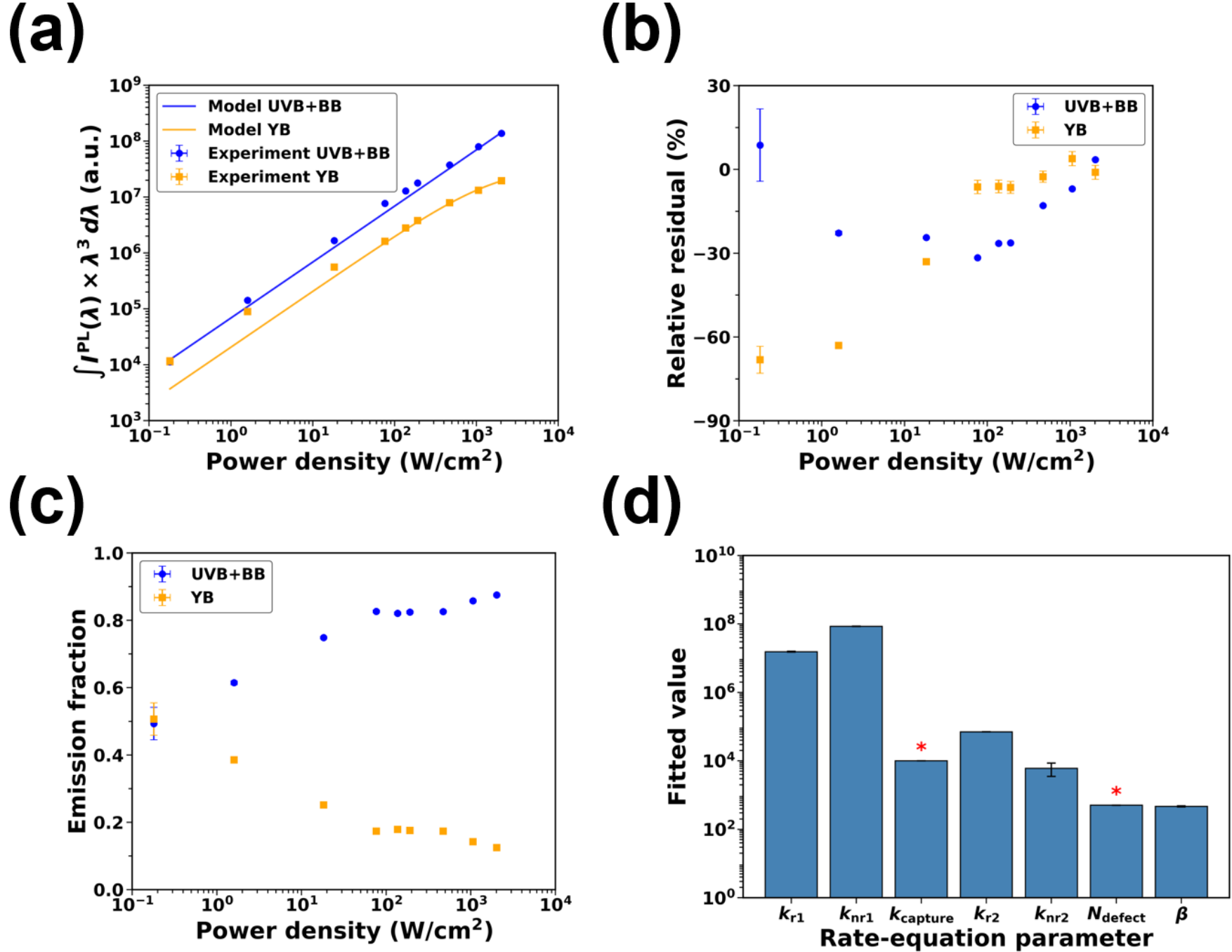}
      \caption{Summary of rate-equation modelling for PL obtained after 1000 °C annealing. (a) Comparison between the emission channels' experimental PL integrated intensities (data points) and the best fit obtained using the rate-equation model (lines) as a function of excitation power density. (b) Relative residuals between the emission channels' experimental and modelled PL integrated intensities. (c) Relative channel contribution of the experimental PL integrated intensities. (d) Mean fitting parameter values and standard deviation obtained from repeated differential evolution runs. An `*' identifies a parameter whose best-fit value lies within 5\% of a search-space boundary.}
      \label{figure_REM_1000}
    \end{figure}

For clarity, Table \ref{table:model parameters} compiles all model parameters alongside their respective operational definitions and constraint boundaries. Within this numerical framework, carrier populations are treated as equivalent dimensionless quantities. This abstraction focuses on accurately delineating the macroscopic trends of dynamic evolution rather than reproducing absolute real-world concentrations, thereby eliminating geometric uncertainties associated with the laser profile and excitation volume. Using data extracted from Figures \ref{figure_PD_PL}(a) and \ref{figure_PD_PL}(b) (after 500 °C and 1000 °C annealing), the results of the rate equation modelling are shown in Figures \ref{figure_REM_500} and \ref{figure_REM_1000}. 

Figures \ref{figure_REM_500}(a) and \ref{figure_REM_1000}(a) show the excitation power density-dependent PL integrated intensities of the (UVB+BB) and YB channels alongside the fit obtained using the rate-equation model. The continuous model curves were generated by independently re-solving the steady-state rate equations at 300 logarithmically spaced power values spanning the experimental range, using the best-fit parameter set obtained from the global optimisation, and do not carry a propagated confidence band reflecting parameter uncertainty. The model reproduces the overall power density dependence shape of both channels across the full excitation range, with the coefficient of determination ($R^2$) for all fitting curves greater than 0.99 (see Supplementary Information: section 6). However, as will be discussed later, high $R^2$ does not imply that all fitted parameters are individually well-constrained.

Figures \ref{figure_REM_500}(b) and \ref{figure_REM_1000}(b) show the relative residuals between the emission channels' experimental and modelled PL integrated intensities as a function of excitation power density. While both channels agree with the model within $\sim$10\% at high excitation power densities (1059.97 $\text{W/cm}^2$), the YB channel shows a systematic underestimation of $-33\%$ to $-81\%$ at low power density (0.18 to 18.30 $\text{W/cm}^2$), reproducible across repeated independent fits. We considered whether this low-power discrepancy might instead be an artefact of the uniform weighting ($w_i=1$) adopted in Equation \ref{equ:data loss}. Increasing $w_i$ for low excitation power density data points was found to improve the low power density residual but degrade the fit at high power densities, indicating a genuine tension that a single set of rate-equation parameters cannot simultaneously resolve across the full excitation range, rather than a weighting artefact. This points to a structural origin: the model assumes a power density-independent carrier capture coefficient $k_{\mathrm{capture}}$, whereas the true capture efficiency in InGaN/GaN MQWs may depend on instantaneous carrier density and defect-state occupation, particularly under weak excitation where defect states are largely unoccupied. A power density-dependent capture model could in principle improve the fit across the full range, but was not pursued here, as it would introduce additional free parameters not warranted by the present dataset ($n=9$).

Figures \ref{figure_REM_500}(c) and \ref{figure_REM_1000}(c) show the excitation power density-dependent fractional contribution of each emission channel to the total PL integrated intensity, computed directly from the experimentally measured integrated intensities of each Gaussian-fitted emission band. Unlike other subplots, it does not depend on the rate-equation parameterisation or on any of the search-space bounds and penalty-window choices. The YB fraction decreases with increasing excitation power density, consistent with progressive saturation of the defect-mediated channel relative to the excitonic (UVB+BB) channel. Annealing to 1000 °C has altered the relative emission contributions of the two channels, with the intersection point shifting from occurring at a moderate power density ($\sim$40 $\text{W/cm}^2$) to a lower power density($\sim$0.18 $\text{W/cm}^2$).

Figures \ref{figure_REM_500}(d) and \ref{figure_REM_1000}(d) show the mean and standard deviation of some important fitted rate-equation parameters across repeated independent global optimisation runs, and the error bar reflects the reproducibility of the optimiser given the data, which is not equivalent to a parameter confidence interval. Asterisks denote parameters whose best-fit value lies within 5\% of a search-space boundary, flagged for the reader's attention rather than indicating a fitting failure. Despite most parameters presenting excellent reproducibility across repeated fits, this does not by itself indicate that a parameter is uniquely well-determined. The values obtained for $k_{\mathrm{capture}}$ and $N_{\mathrm{d}}$ are highly reproducible yet consistently sit at the lower edge of their search range, indicating boundary-truncated rather than data-determined solutions. $k_{\mathrm{nr2}}$ shows the largest run-to-run variation without being pinned at either boundary, indicating a shallow rather than sharply bounded region of the loss landscape, consistent with relaxation from the YB channel being strongly radiatively dominated ($k_{\mathrm{r2}}\gg k_{\mathrm{nr2}}$). $k_{\mathrm{r1}}$ and $\beta$ are consistently reproduced with little variation, but are found to be linked by an exact multiplicative degeneracy such that their individual values depend on the prior bound imposed on $k_{\mathrm{capture}}$, rather than being uniquely determined by the data. Only $k_{\mathrm{nr1}}$ and $k_{\mathrm{r2}}$ are both strongly reproducible and free of these caveats. We therefore report these fitted parameters, together with their derived quantities $\tau_{1, 2}$ and $\eta_{\mathrm{UVB+BB, YB}}$ as bound-dependent estimates, and they should not be over-interpreted as precisely physical quantities.

The absolute values of the seven coupled fitting parameters should not be treated as uniquely determined by the limited number of independent datasets. However, their relative trends are of interest for cross-condition comparisons, given that identical parameter bounds and regularisation are strictly applied throughout. As depicted in Figures \ref{figure_REM_500}(d) and \ref{figure_REM_1000}(d), the extracted non-radiative recombination coefficient for the (UVB+BB) channel ($k_{nr1}$) exhibits remarkable constancy upon increasing the annealing temperature from 500 °C to 1000 °C, consistently yielding a value of the order of $\sim10^{8}\text{ s}^{-1}$. Concurrently, the equivalent carrier lifetime $\tau_1$ yields a consistent value of $\sim$9.9 ns (Table \ref{table:equivalent lifetime}). This invariant nature provides macroscopic, quantitative evidence of the pristine structural integrity of the InGaN active regions, verifying that the RTA process combined with the AlN capping layer successfully circumvents catastrophic interdiffusion or macro-scale phase separation. Interestingly, under high-excitation power density, the radiative coefficient $k_{r1}$ exhibits a visible upward shift from $\sim10^{6.7}\text{ s}^{-1}$ to $\sim10^{7.2}\text{ s}^{-1}$, driving the equivalent efficiency $\eta_{\mathrm{(UVB+BB)}}$ from 5.48\% up to 14.82\%. Considering the YB channel, while the non-radiative parameter $k_{nr2}$ climbs from $\sim10^{3.3}\text{ s}^{-1}$ to $\sim10^{3.9}\text{ s}^{-1}$, the radiative parameter $k_{r2}$ reduces from $\sim10^{5.3}\text{ s}^{-1}$ to $\sim10^{4.8}\text{ s}^{-1}$ upon higher temperature annealing, suggesting proliferation of non-radiative centres and reconstruction or passivation of the deep level defect complexes.

The transformation of the crystalline environment with annealing is evidenced with far greater sensitivity under the analysis of the lower excitation power densities used in the unfocused laser configuration (Supporting Information: section 6). Here, the lower excited carrier density avoids defect saturation of traps and acts as a sensitive probe for carrier dynamics. Upon 1000 °C annealing, a strong thermal effect within the GaN matrix is unambiguously revealed by the YB channel. Its non-radiative rate ($k_{nr2}$) experiences a substantial boost from $\sim10^{3.5}\text{ s}^{-1}$ (500 °C) to $\sim10^{4.7}\text{ s}^{-1}$ (1000 °C). This signifies an increased proliferation of thermally propagated non-radiative defect networks that aggressively capture photo-generated carriers. Meanwhile, the YB channel's radiative parameter $k_{r2}$ experiences a reduction from $\sim10^{5.3}\text{ s}^{-1}$ to $\sim10^{5.0}\text{ s}^{-1}$. In sharp contrast, a highly anomalous, counter-intuitive trend is observed for the (UVB+BB) channel. Specifically, the radiative recombination rate $k_{r1}$ under the unfocused scheme experiences a dramatic, order-of-magnitude increase from $\sim10^{6.3}\text{ s}^{-1}$ at 500 °C to $\sim10^{7.8}\text{ s}^{-1}$ at 1000 °C, while its non-radiative counterpart $k_{nr1}$ drops from $\sim10^{8}\text{ s}^{-1}$ to $\sim10^{7.5}\text{ s}^{-1}$. 

This behaviour provides direct evidence of a carrier localisation shielding effect. Annealing at 1000 °C, the heavy thermal budget stochastically drives and stabilises nanoscopic indium-rich compositional fluctuations into highly robust potential minima. The consequences of this thermodynamic evolution are vividly reflected in the derived lifetimes and radiative efficiencies as defined in Equations \ref{equ:emission lifetime} and \ref{equ:emission efficiency}. A comprehensive summary regarding these deduced values can be found in Table \ref{table:equivalent lifetime}. Under low power density excitation, the passivation of the deep level defect complexes and the proliferation of thermally propagated defect networks cause the radiative efficiency of the YB ($\eta_{\mathrm{YB}}$) to collapse drastically from 98.55\% down to 67.17\%. Remarkably, under the same low power density excitation condition, the efficiency of the (UVB+BB) channel ($\eta_{\mathrm{(UVB+BB)}}$) undergoes a $\sim$30-fold increase, climbing from a negligible 2.18\% (500 °C) to 66.02\% (1000 °C). Even though the equivalent lifetime $\tau_1$ is slightly increased to 10.34 ns, the increase in $k_{r1}$ guarantees that carriers are rapidly captured and tightly confined within these nanoscopic clusters upon generation. This rapid spatial separation effectively prevents carriers from migrating outwards into the defect-enriched bulk networks, establishing the consolidated localisation centres as a `thermodynamic shield' that sustains the stable radiative output of the MQW.

    \begin{table}[!ht]
        \centering
        \caption{Summary of the emission channels' lifetimes and radiative efficiencies obtained from modelling under different conditions.}
        \begin{tabularx}{\textwidth}{X X X X X}
        \hline
            Conditions &(UVB+BB) lifetime (ns) &(UVB+BB) efficiency (\%) & YB lifetime ($\mu s$) & YB efficiency (\%)\\ \hline
            Post-annealed at 500 °C, focused & 9.88 & 5.48 & 4.94 & 98.93\\ 
            Post-annealed at 1000 °C, focused & 9.89 & 14.82 & 12.71 & 89.92\\
            Post-annealed at 500 °C, unfocused & 9.97 & 2.18 & 4.93 & 98.55\\
            Post-annealed at 1000 °C, unfocused & 10.34 & 66.02 & 6.49 & 67.17\\
            \hline
        \end{tabularx}
        \label{table:equivalent lifetime}
    \end{table}

\subsection{Chromaticity coordinate analysis}
\begin{figure}[htbp]
      \centering
      \includegraphics[width=1\textwidth]{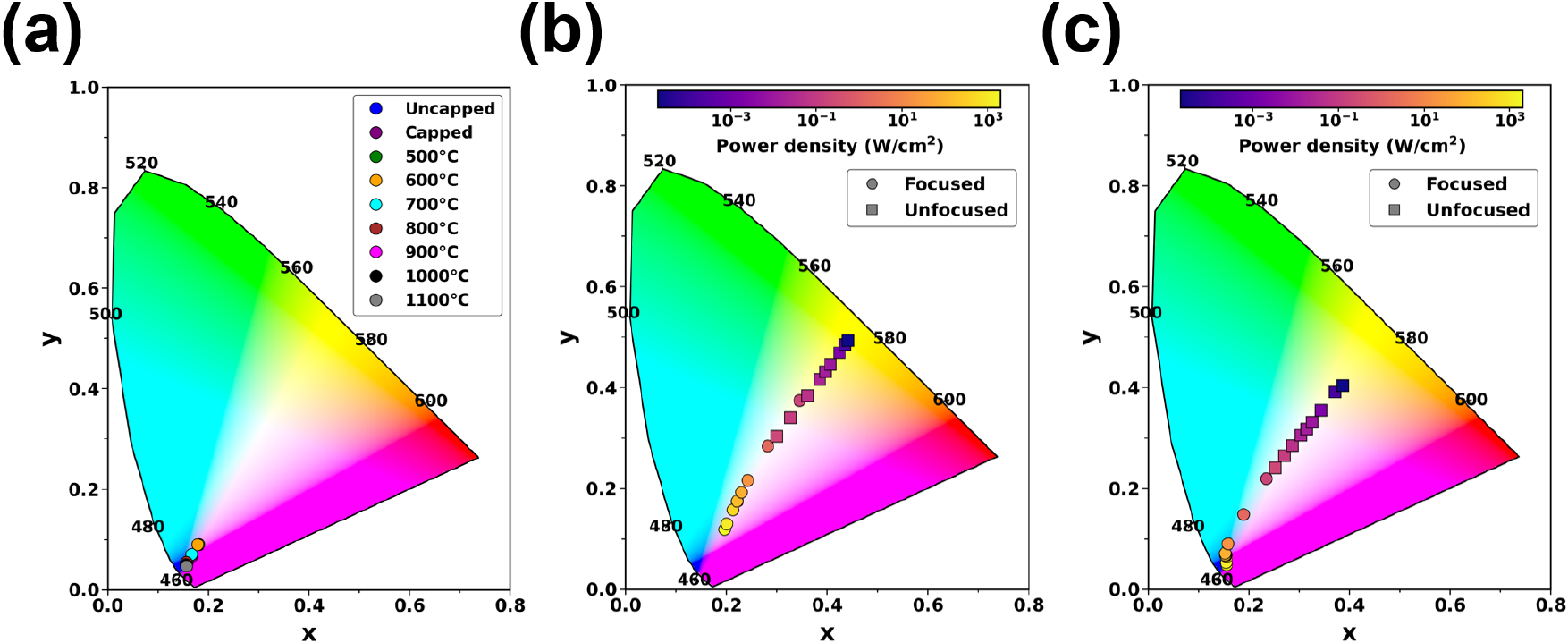}
      \caption{CIE chromaticity diagrams obtained using (a) 2021.27 $\text{W/cm}^2$ 325 nm excitation as a function of annealing temperature, and as a function of excitation power density following annealing at (b) 500 °C and (c) 1000 °C.}
      \label{figure_CIE}
    \end{figure}

To analyse the colourimetric properties of the sample emission, the room temperature PL spectra shown in Figures \ref{figure_RT_PL}(a), \ref{figure_PD_PL}(a), and \ref{figure_PD_PL}(b) were mapped onto the CIE chromaticity diagrams, as depicted in Figures \ref{figure_CIE}(a) to \ref{figure_CIE}(c). 

As presented in Figure \ref{figure_CIE}(a), the colour coordinates of the sample emission show a small variation under the highest excitation power density used, confined to the blue region, even after thermal annealing up to 1100 °C. This macroscopically indicates that the MQW remains structurally sound, resisting catastrophic thermal dissolution under such thermal budgets.

As discussed above, reducing the excitation power density changes the relative contribution of the BB and YB to the total emission. This results in the ability to tune the colour coordinate as shown in Figure \ref{figure_CIE}(b) and (c) for the 500 °C and 1000 °C annealed samples respectively. With reducing excitation power density the colour is altered from close to the blue perimeter towards the centre of the diagram, though with a reduction in the overall emission intensity. The purest blue emission is obtained from the sample post 1000 °C annealing and when using the highest excitation power density available yielding x and y coordinates of 0.156 and 0.049 respectively. Under the lowest excitation densities used (plotted using square data point symbols) the dominant YB emission shifts the colour towards the saturation boundary in the yellow spectral region. In this case annealing at 500 °C (1000 °C) and use of the lowest excitation power density yields x,y coordinates of 0.441, 0.494 (0.387, 0.404).

\subsection{Time-of-flight secondary ion mass spectrometry analysis}
\begin{figure}[htbp]
      \centering
      \includegraphics[width=1\textwidth]{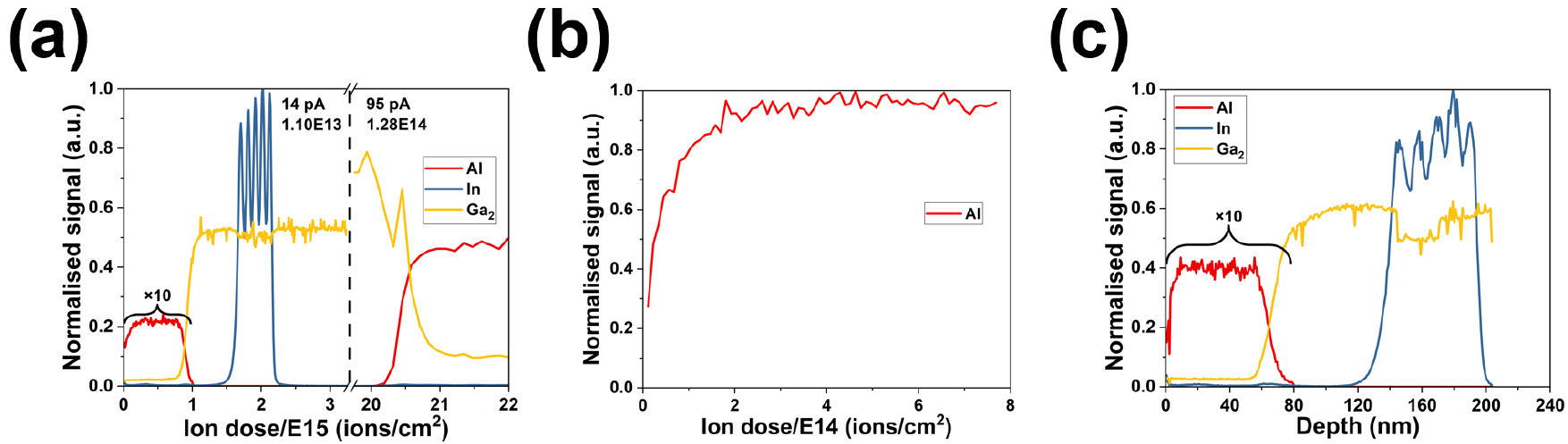}
      \caption{The ToF-SIMS profile of the AlN/GaN/MQW/GaN/Al$_2$O$_3$ sample. (a) Obtained using an ion current of 14 pA in the initial phase and continued using an increased ion current of 95 pA to reach the Al$_2$O$_3$ substrate. (b) Shallow profile used to obtain the AlN etch rate using a current of 8 pA. (c) Profile similar to (a) but obtained using a current of 8 pA. Initial signals of the Al element in (a) and (c) have been enlarged by a factor of 10 for clarity.}
      \label{figure_ToF_SIMS}
    \end{figure}

To elucidate the composition and evaluate the structural integrity of the MQW following the serial thermal anneals up to 1100 °C, ToF-SIMS depth profiling was conducted. A total of three sites were profiled for data analysis. Figure \ref{figure_ToF_SIMS}(a) illustrates the ToF-SIMS result of a deep etch (site 1) used to sputter through the entire sample structure spanning from the outermost surface layer down to the underlying sapphire substrate. We note that the ion flux was increased from 1.10E13 to 1.28E14 ions/cm$^2$ following an accumulative dose of 3.3E15 ions/cm$^2$ to enable the sapphire substrate to be reached within a reasonable time (indicated by the dashed line and axis break in the figure). The two further ToF-SIMS profiles taken (sites 2 and 3) were restricted in depth in order to enable estimation of the etching rates of the AlN cap and underlying GaN. The profiles of these are presented in Figure \ref{figure_ToF_SIMS}(b) and \ref{figure_ToF_SIMS}(c). To prevent detector saturation, the gallium dimer (Ga$_2$) signal was selected for plotting due to the lower prevalence of this species. 

Figure \ref{figure_ToF_SIMS}(a) presents a continuous ToF-SIMS profile through the sample. The Al signal resulting from the AlN thin film cap is first observed prior to the Ga$_2$ signal emerging as a result of sputtering the GaN layer. There then follows a strong  In signal exhibiting highly periodic, sharp oscillatory peaks, revealing the presence of buried InGaN MQWs. These appear to be well-defined and preserved without catastrophic interdiffusion. It is observed that the Al profile shows no diffusion of Al into the GaN and MQW region, confirming that the AlN capping layer functions as an excellent out-diffusion barrier without introducing self-contamination into the active layers. Due to the sample thickness, the ToF-SIMS sputtering dose was elevated during the analysis to increase the sample etching rate after passing through the MQW structure in order to reach the underlying substrate. This is shown after crossing the dashed line in Figure \ref{figure_ToF_SIMS}(a), accompanied by the abrupt increase in the Ga$_2$ signal. After accumulating a total dose of around 2E16 ions/cm$^{2}$, the main structure is finally etched through, and an Al signal from the sapphire substrate appears. Overall, the ToF-SIMS profile provides structural evidence of the thermal stability of the MQW structure.

As the secondary ion sputtering rate is composition-dependent, further shallow ToF-SIMS profiles were taken to obtain the sputtering rates of the AlN capping layer and the GaN matrix. These were then profiled using AFM to obtain etch rates. Figure \ref{figure_ToF_SIMS}(b) presents the Al signal from site 2, where the sputtering was stopped before reaching the AlN/GaN interface. The crater thickness was measured by AFM to be approximately 31.62 ±1.22 nm (see Supporting Information, section 7), yielding an etch rate of 4.59 nm per 1E14 ions/cm$^2$. The slow rise of the Al signal in Figure \ref{figure_ToF_SIMS}(b) is a surface transient effect induced by the progressive implantation of primary ions and the un-equilibrated secondary ion yields near the immediate surface. To account for this the initial transient region, defined as the segment where the signal intensity is below 75\% of the global plateau maximum, was excluded from the etch rate estimation.

ToF-SIMS analysis of site 3 was performed, stopping after passing through the MQW structure, Figure \ref{figure_ToF_SIMS}(c). Utilising the etch rate obtained for AlN from site 2, the etch rate for GaN was obtained following AFM measurement of the crater depth profile ($\sim$201.43 ±10.22 nm, see Supporting Information, section 7) as 7.76 nm per 1E14 ions/cm$^2$. The MQW structure appears less resolved in this case which is an experimental artefact resulting from the knock-on effect due to sputter-induced mixing. Figure \ref{figure_ToF_SIMS}(c) is plotted using a depth scale based on these obtained AlN and GaN sputter rates. The five InGaN/GaN QWs were identified by globally fitting the In SIMS depth profile with a sum of five Gaussian peaks and a shared, continuous polynomial background. The fitted peak centres yielded an average period of 11.68 ±0.54 nm, with the quoted uncertainty given by the standard error of the mean across the four adjacent periods, reflecting the layer-to-layer variation in growth period rather than the (smaller) uncertainty on any individual peak centre from the fit covariance matrix. The first QW was located 91.01 ±0.22 nm beneath the AlN/GaN interface, where the uncertainty is the fit-covariance-based standard error on that single peak's centre.

\section{Conclusion}

In summary, we have explored the thermal stability and carrier recombination kinetics of an InGaN/GaN MQW structure through room-temperature photonic measurements. The preservation of the MQW following multiple anneals up to 1100 °C was verified via ToF-SIMS analysis. The main impact of thermal treatment on the emission characteristics of the MQW is to regulate its luminescence intensity, as would be expected due to combined effects of heat-induced damage and structural recovery. Thermal annealing at temperatures from 500 °C to 1000 °C/1100 °C is observed to result in very limited changes in the MQW band's peak position, above-bandgap absorption behaviour, carrier recombination mechanism, calculated equivalent carrier lifetime, and emission colour, as suggested by the steady-state PL spectra, PLE profiles, excitation power density-dependent measurements, rate equation modelling results, and CIE diagrams, respectively. These findings collectively demonstrate the excellent thermal robustness of the MQW optical properties within the evaluated temperature window.

\section{Acknowledgements}

This work was funded by EPSRC grant EP/V001914/1, and by capital investment by the University of Manchester. Q-S.L thanks the China Scholarship Council for financial support. The authors thank R Oliver (University of Cambridge) for providing the sample studied in this work, and D Binks (University of Manchester) and S Church (University of Salford) for useful discussions. The authors would also like to thank C Smith (University of Manchester) and J Jacobs (University of Manchester) for their valuable contributions and assistance during the early stages of the experimental work.

\section{Author contributions statement}

Q-S.L, S.S, S.G, M.C, C.S, and J.J conducted all experimental work. Q-S.L, S.S, M.C and R.C performed data analysis. N.L and R.C supervised the work. All authors contributed to the writing of the manuscript.

\section{Competing interests}

The author(s) declare no competing interests.

\bibliographystyle{unsrtnat}
\bibliography{references}  %%% Uncomment this line and comment out the ``thebibliography'' section below to use the external .bib file (using bibtex).

\clearpage
\begin{center}
    \Large
    \textbf{Supplementary Information}
    \\[20pt]
    \normalsize 
\end{center}
\setcounter{figure}{0}
\setcounter{table}{0}
\setcounter{section}{0}
\setcounter{equation}{0}
\renewcommand*{\thefigure}{S\arabic{figure}}
\renewcommand*{\thetable}{S\arabic{table}}
\renewcommand*{\theequation}{S\arabic{equation}}

\section{Deposition of the AlN thin film}
\label{SectionS1}
\begin{figure}[htbp]
      \centering
      \includegraphics[width=0.8\textwidth]{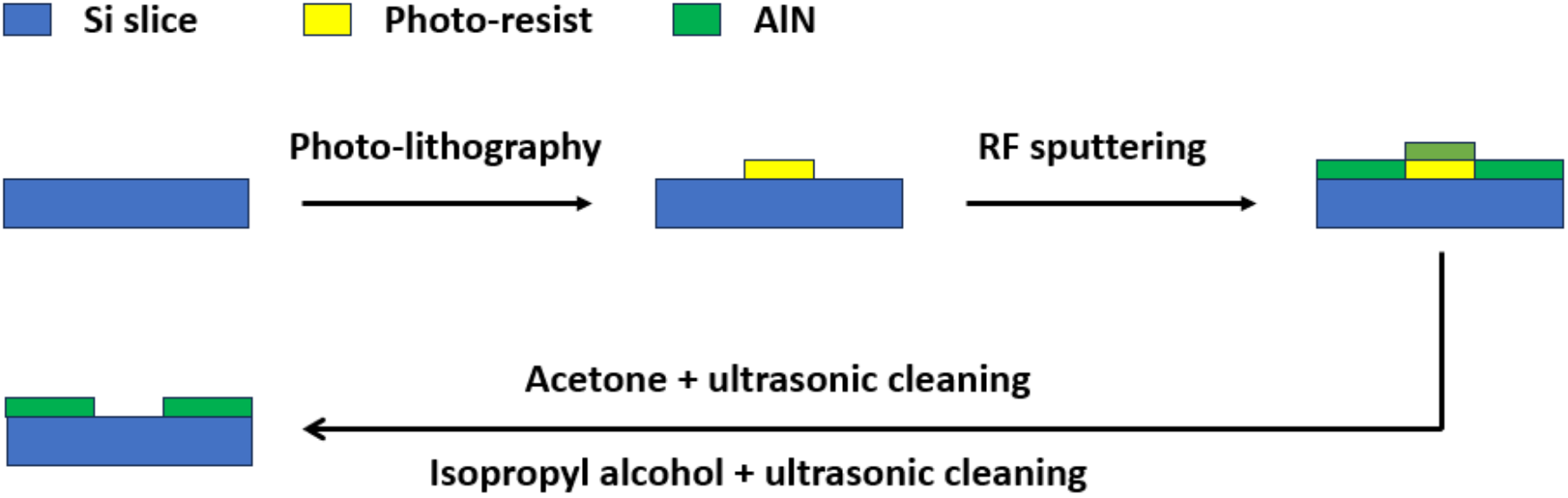}
      \caption{Lateral view schematic illustration of the formation of sharp AlN/Si edges.}
      \label{figureS_deposition}
    \end{figure}

As full surface coverage is required for the capping layer, measurement of its thickness without exposing the underlying GaN is a challenge. As such, a silicon wafer was used as an alternative template for enabling the thickness measurement of the AlN capping layer. Photolithography and lift-off were used to create a sharp boundary between the silicon and AlN for measurement. Following deposition of the photoresist and patterning AlN was deposited using RF sputtering. Lift-off was performed by soaking in acetone for several minutes, accompanied by ultrasonication, followed by repeating the process in isopropyl alcohol for several minutes. This provided an AlN thin film with stepped edges enabling film thickness measurement via atomic force microscopy (AFM). For clarity, the above process is illustrated in Figure \ref{figureS_deposition}. 

    \begin{figure}[htbp]
      \centering
      \includegraphics[width=1\textwidth]{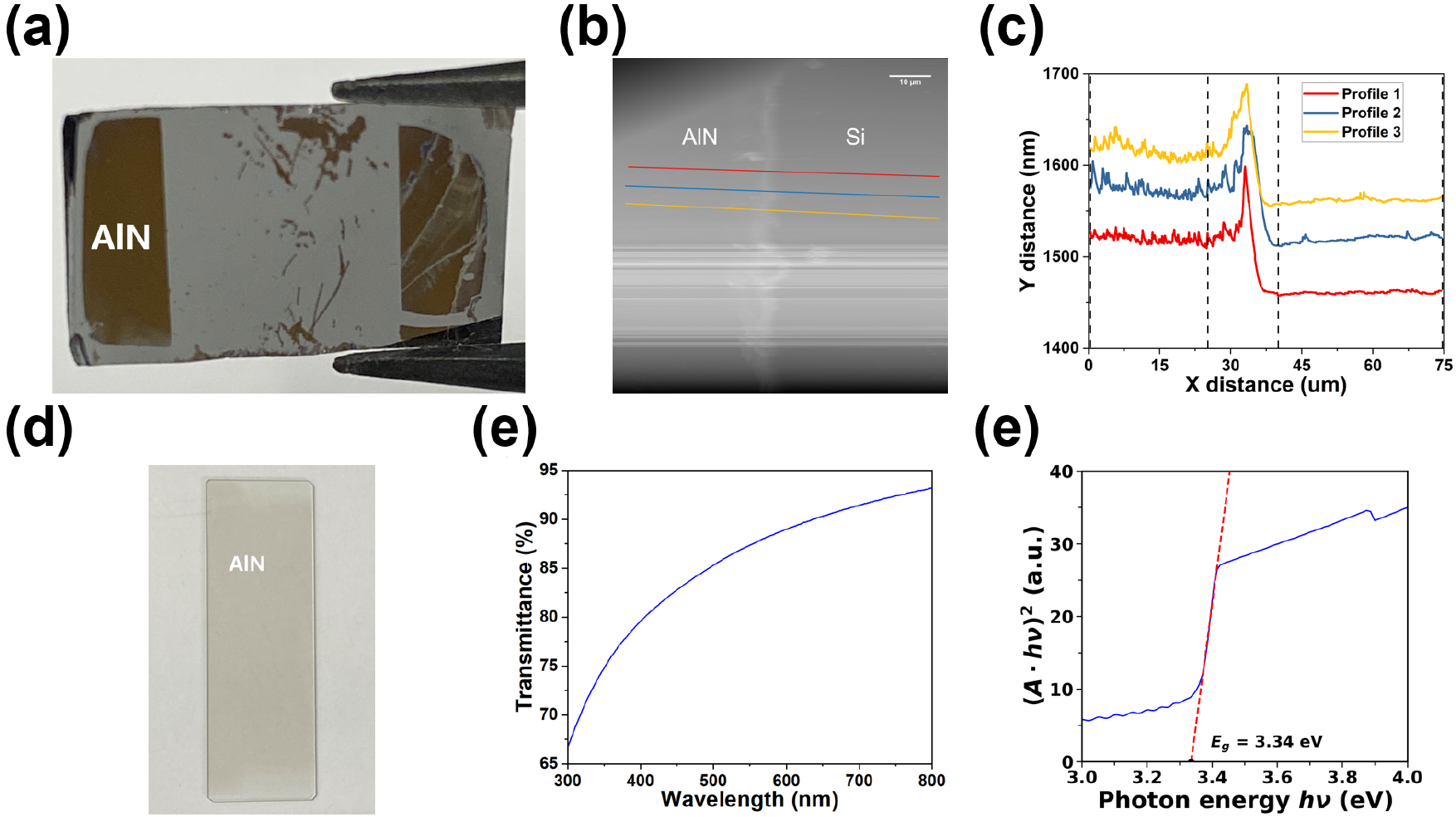}
      \caption{(a) The physical image of the silicon slice after AlN sputtering. (b) The AFM topography of the AlN/Si boundary with three cross-sectional lines. (c) Corresponding depth profiles colour-coded to match the lines in (b). (d) The physical image of the glass slide after AlN sputtering. (e) The transmittance spectrum of the as-deposited glass slide displayed in (d). (f) Tauc plot derived from the UV-Vis absorbance spectrum of the sample.}
      \label{figureS_AlN_thickness}
    \end{figure} 

The AlN deposition was performed simultaneously onto the as-received GaN sample, a piece of Si wafer, and a bare glass slide adjacently mounted in the chamber. Figures \ref{figureS_AlN_thickness}(a) and \ref{figureS_AlN_thickness}(d) show the deposited AlN film on the silicon (post lift-off) and the glass slide respectively. To determine the film thickness (using the step height), three independent line profiles were extracted across the AlN/Si boundary, as colour-coded in Figure\ref{figureS_AlN_thickness}(b) and Figure \ref{figureS_AlN_thickness}(c). For each profile, the step height was calculated by taking the difference between the average y values within two distinct baseline regions (one on the AlN layer and the other on the Si substrate), which are demarcated by the vertical dashed lines in Figure \ref{figureS_AlN_thickness}(c). By averaging the height differences calculated from these three profiles, a mean step height of 55.67 ±1.67 nm was obtained, with the error representing the standard error of the mean. The transmittance spectrum of the glass slide was also measured with a Perkin-Elmer Lambda 1050 UV/Vis/NIR spectrometer, Figure \ref{figureS_AlN_thickness}(e), presenting high transmittance across the visible spectral region. Following the highest temperature anneal (1100 °C), the same instrument was used to determine the sample's UV-Vis absorption spectrum which was used to estimate the optical bandgap $E_g$ via the Tauc relation:

 \begin{equation}     
     (A \cdot h\nu)^2 = C(h\nu - E_g),     
     \label{eq:tauc} 
 \end{equation} 
 
where $A$ is the measured absorbance, $h\nu$ is the photon energy, and $C$ is a constant. The exponent of 2 was adopted assuming a direct allowed transition, consistent with the direct bandgap nature of wurtzite GaN. As shown in Figure \ref{figureS_AlN_thickness}(f), the linear region of the $(A \cdot h\nu)^2$ vs. $h\nu$ plot near the absorption edge was fit and extrapolated to $(A \cdot h\nu)^2 = 0$, yielding an optical bandgap of $E_g = 3.34$ eV, in good agreement with the accepted value for wurtzite GaN at room temperature ($\sim$3.4 eV).

\section{Correction curve for the spectral response}
\label{SectionS2}
\begin{figure}[htbp]
      \centering
      \includegraphics[width=0.8\textwidth]{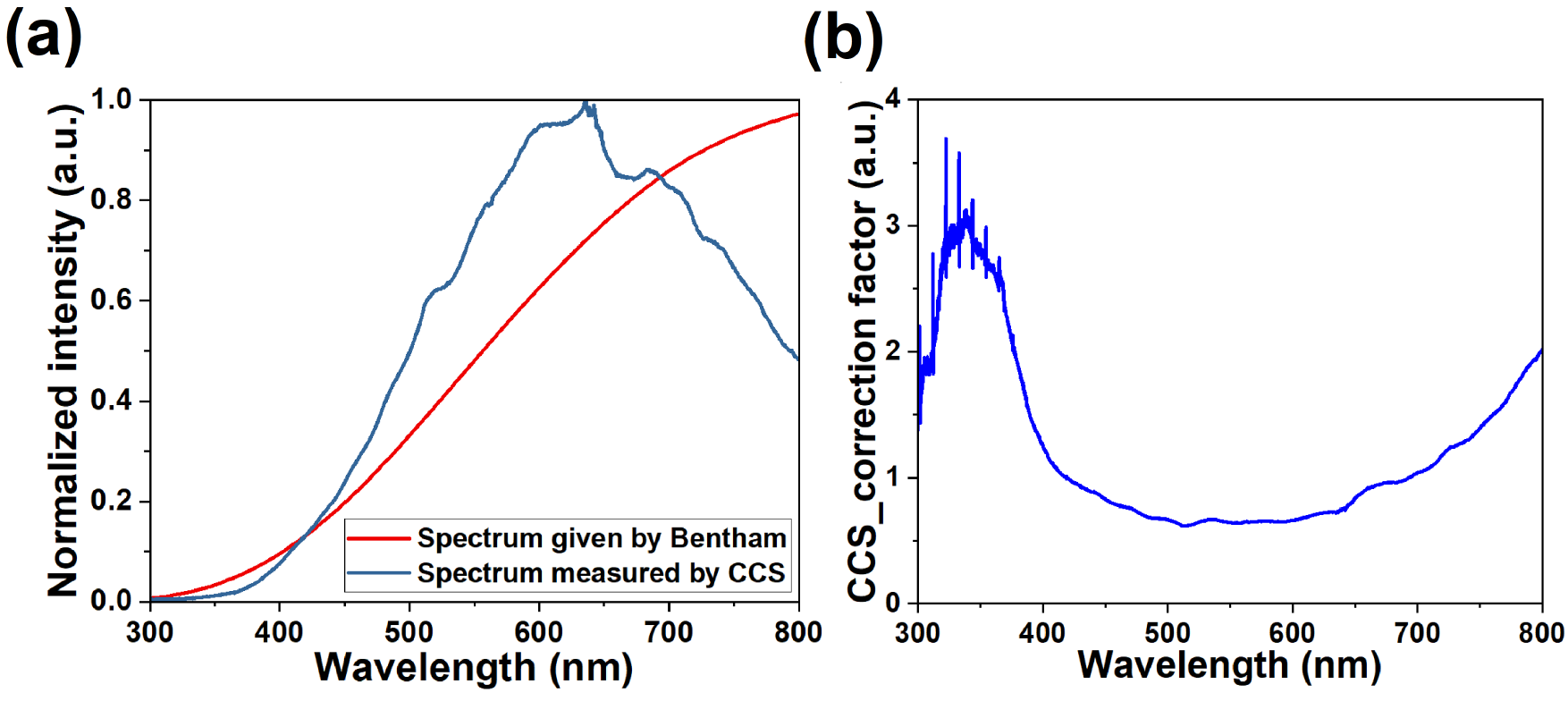}
      \caption{(a) Normalised spectra of IL1 radiation given by Bentham and measured by the CCS200 spectrometer. (b) The correction factor curve of the CCS200 spectrometer.}
      \label{figureS_correction}
    \end{figure}

Figure \ref{figureS_correction}(a) exhibits two normalised spectra of the Bentham IL1 halogen source radiation, one provided by the manufacturer and the other acquired using the CCS200 spectrometer. A correction factor curve (Figure \ref{figureS_correction}(b)) that includes fibre coupling and transmission loss can be derived by setting the Bentham spectrum as the numerator and the measured spectrum as the denominator, respectively.

\section{Three emission bands' integrated intensity evolution on a linear scale}
\label{SectionS3}

\begin{figure}[htbp]
      \centering
      \includegraphics[width=1\textwidth]{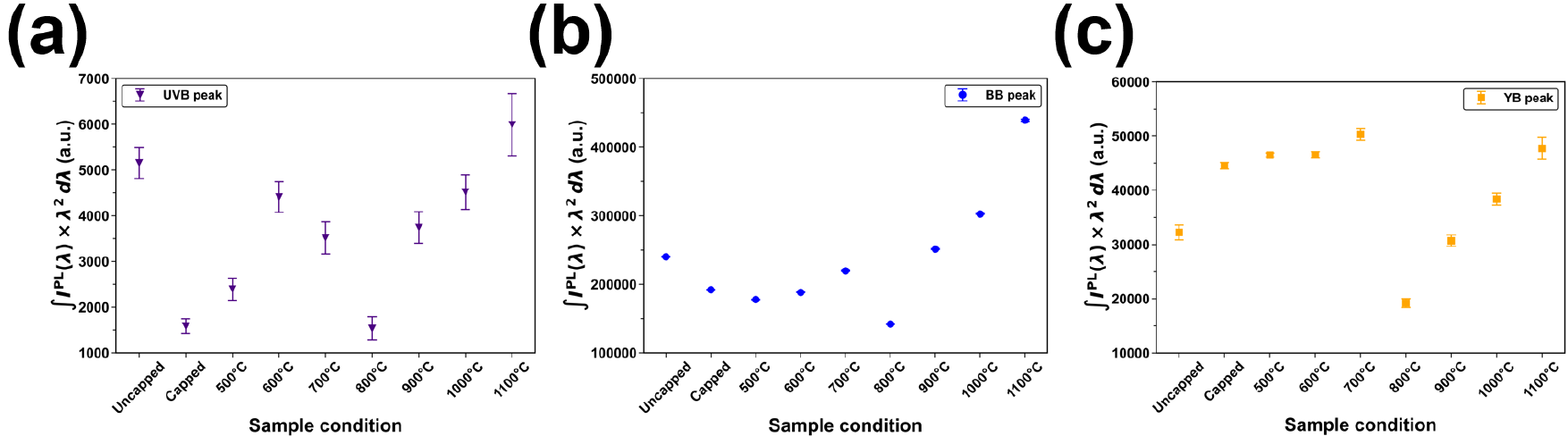}
      \caption{The integrated intensity evolution of (a) UVB, (b) BB, and (c) YB emission.}
      \label{figureS_RT_PL_linear}
    \end{figure}

Figure \ref{figureS_RT_PL_linear} replots the integrated intensity evolution of each emission band shown in Figure \ref{figure_RT_PL}(d) using a linear scale.

\section{Time-resolved emission decay fitted via bi-exponential model}
\label{SectionS4}

\begin{figure}[htbp]
      \centering
      \includegraphics[width=1\textwidth]{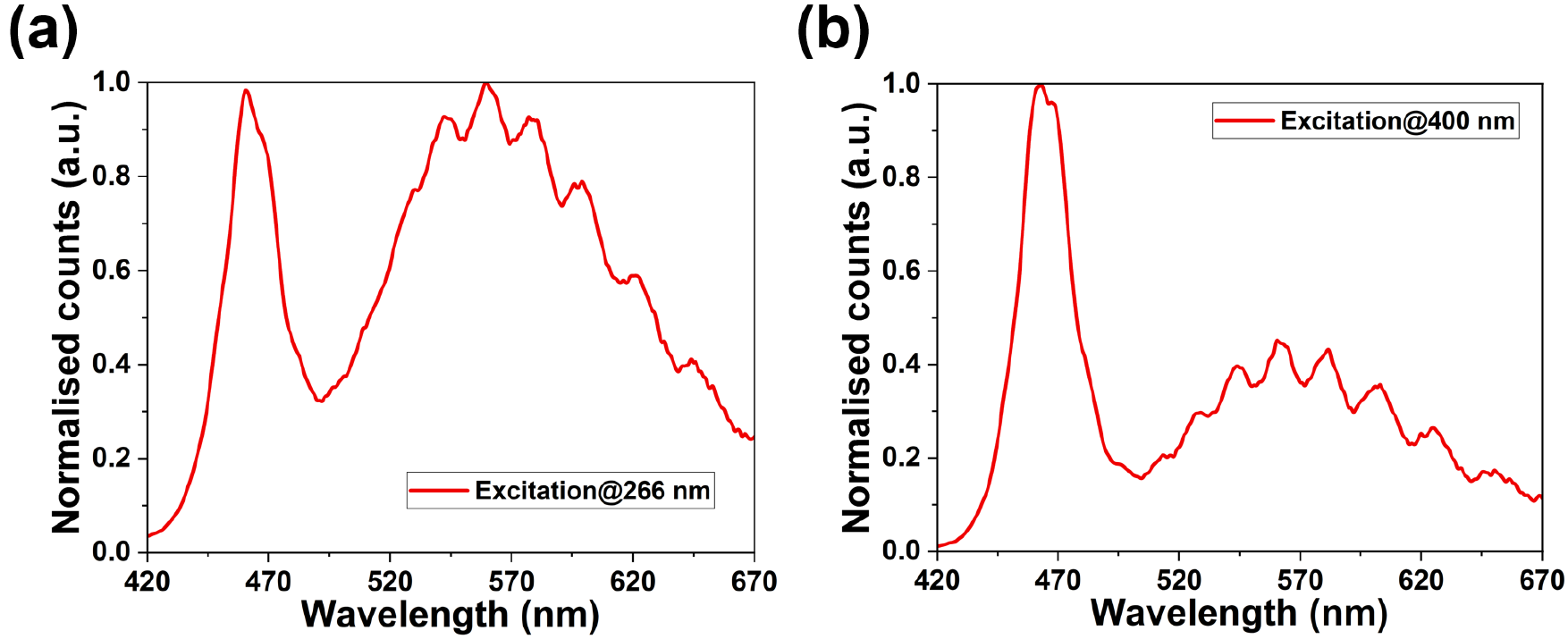}
      \caption{Normalised steady-state PL obtained using a low-intensity excitation wavelength of (a) 266 nm ($\sim$0.19 $\text{mW/cm}^2$) and (b) 400 nm ($\sim$2.26 $\text{mW/cm}^2$).}
      \label{figureS_emission}
    \end{figure}

Figure \ref{figureS_emission} presents the normalised steady-state PL spectra of the sample at excitation wavelengths of 266 nm and 400 nm. Measurements were performed using the same experimental set-up used for the PL excitation (PLE) spectra. 

A bi-exponential model was used to fit the decay curves (see Equation \ref{equ:lifetime fitting}), where $I(t)$ represents the photon counts at time $t$; $A_{1}$ and $A_{2}$ are amplitude scaling factors; $\tau_{fast}$ and $\tau_{slow}$ denote the intrinsic PL lifetimes of an initial fast component and a second slow component, respectively; B is a free parameter accounting for the constant background noise (e.g., detector dark counts and ambient stray light). The comprehensive carrier dynamics were quantified using the intensity-weighted average lifetime ($\tau_{\text{avg}}$) as defined in Equation \ref{equ:lifetime fitting}. Fitting was performed by minimising the Poisson deviance statistic as described in the main text.

    \begin{equation}
     I(t) = A_1 \cdot \exp\left(-\frac{t}{\tau_{fast}}\right) + A_2 \cdot \exp\left(-\frac{t}{\tau_{slow}}\right) + B, \qquad
     \tau_{\text{avg}} = \frac{A_1 \tau_{fast}^2 + A_2 \tau_{slow}^2}{A_1 \tau_{fast} + A_2 \tau_{slow}}.
    \label{equ:lifetime fitting}
    \end{equation}

    \begin{figure}[htbp]
      \centering
      \includegraphics[width=1\textwidth]{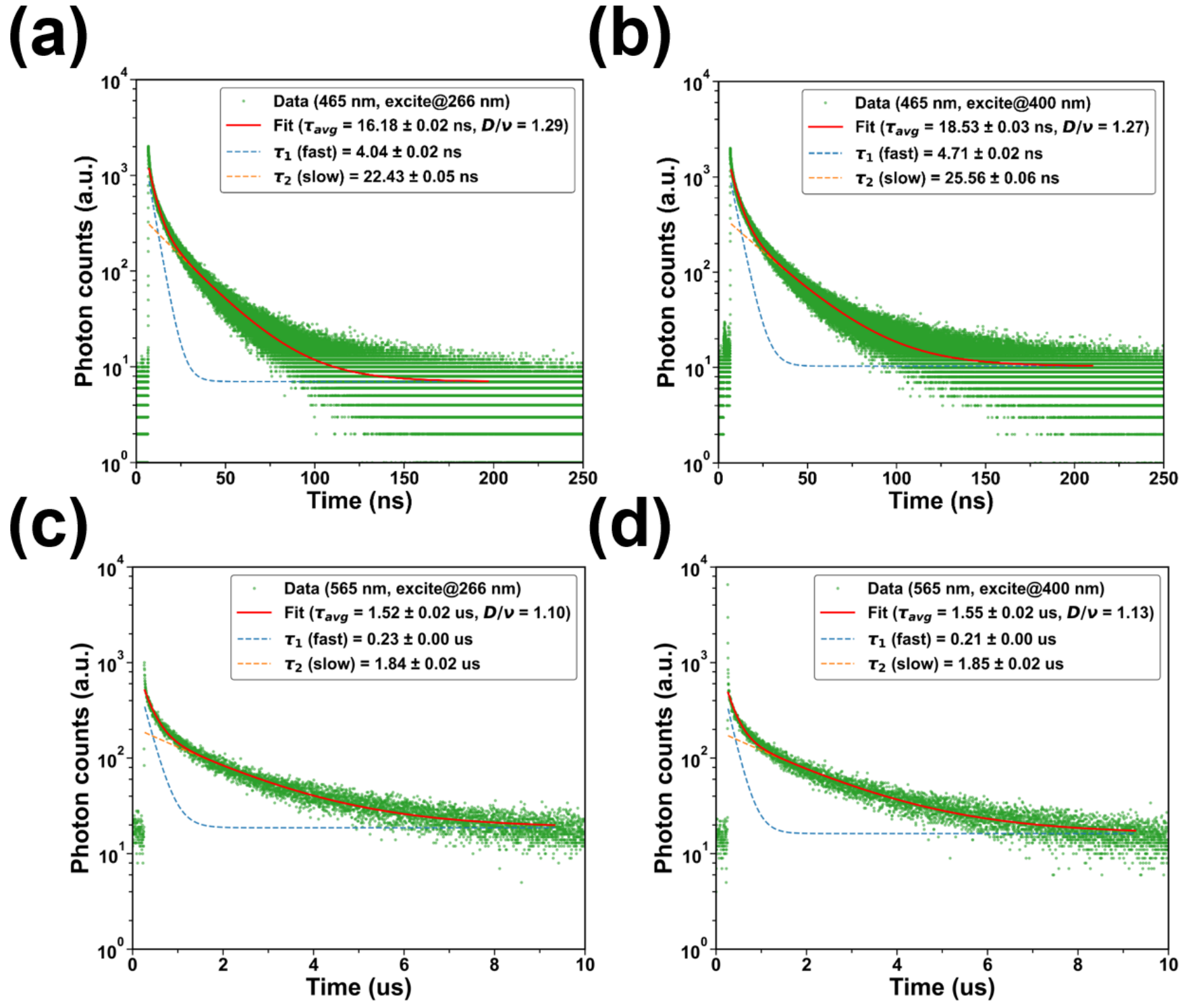}
      \caption{Fitted with the bi-exponential model. PL decay curve following annealing at 1100 °C of the BB ($\lambda_{\text{em}} = 465\text{ nm}$) PL at an excitation wavelength of (a) 266 nm and (b) 400 nm, and the YB ($\lambda_{\text{em}} = 565\text{ nm}$) PL at an excitation wavelength of (c) 266 nm and (d) 400 nm.}
      \label{figureS_lifetime}
    \end{figure} 

Comparison immediately validates the superiority of the stretched exponential model. For the 465 nm BB emission (Figures \ref{figureS_lifetime}(a) and \ref{figureS_lifetime}(b)), although the calculated average lifetimes have similar values ($\tau_{\text{avg}}$ = 16.18 ±0.02 ns for 266 nm, $\tau_{\text{avg}}$ = 18.53 ±0.03 ns for 400 nm), the bi-exponential fitting yields a noticeably compromised goodness-of-fit metric ($D/\nu = 1.29$ for 266 nm, $D/\nu = 1.27$ for 400 nm), whereas the stretched exponential model optimises the metric down to near-ideal unity ($D/\nu = 1.03$ and $1.04$, respectively). This mathematical trend is even more pronounced for the 565 nm YB band (Figures \ref{figureS_lifetime}(c) and \ref{figureS_lifetime}(d)), where the bi-exponential model forces the continuous relaxation spectrum into two discrete fast and slow components ($\tau_{fast} \sim 0.22\ \mu\text{s}$, $\tau_{slow} \sim 1.85\ \mu\text{s}$). Consequently, the bi-exponential average lifetimes yield a deceptively flat response ($\tau_{\text{avg}}$ = 1.52 ±0.02 \textmu s under 266 nm versus $\tau_{\text{avg}}$ = 1.55 ±0.02 \textmu s under 400 nm), creating an illusion of absolute kinetic insensitivity to the excitation wavelengths.

\section{Low excitation power density-dependent measurements using the unfocused laser configuration}
\label{SectionS5}
\begin{figure}[htbp]
      \centering
      \includegraphics[width=1\textwidth]{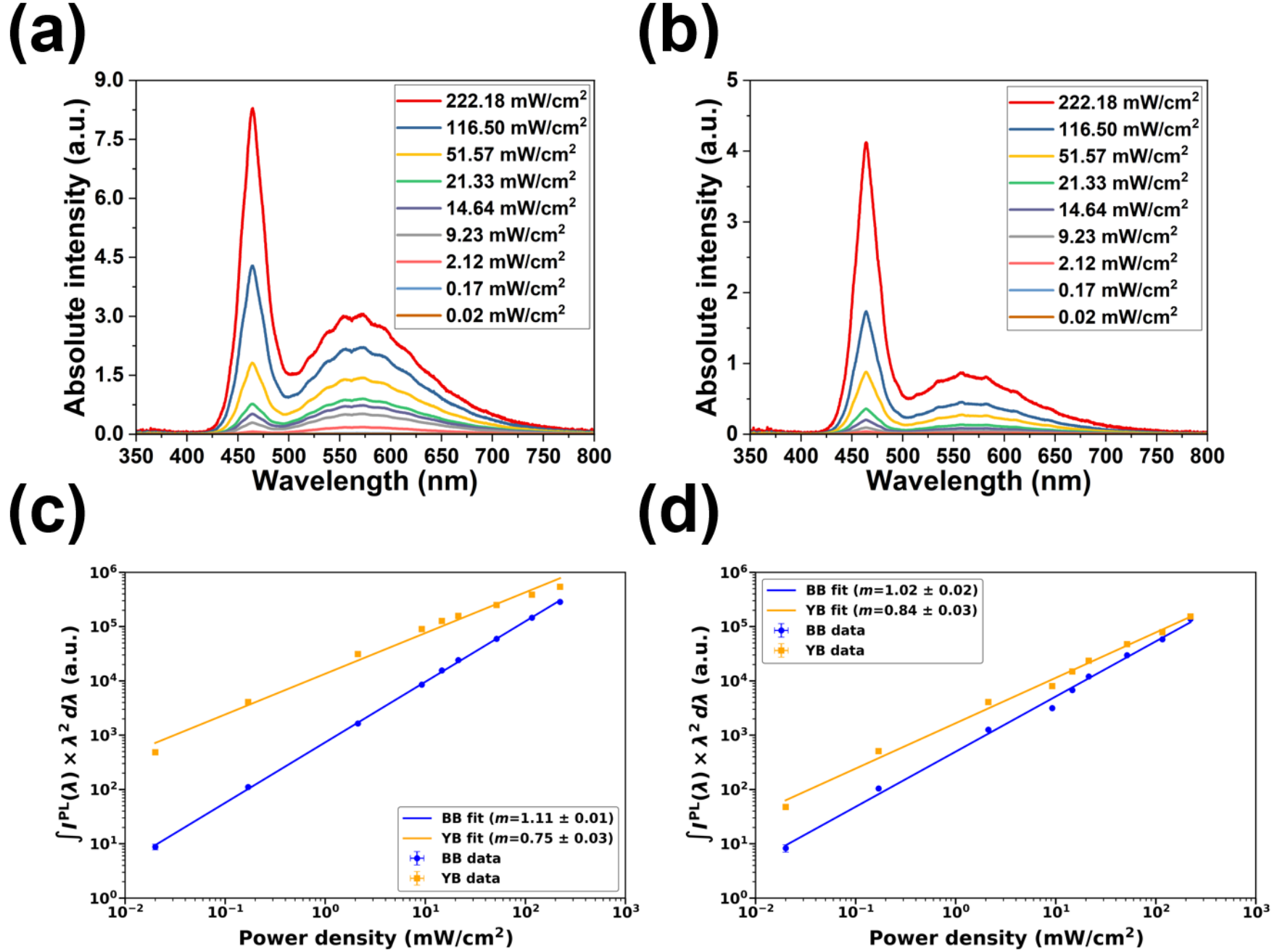}
      \caption{Excitation power density dependence of 325 nm unfocused laser-excited PL obtained after (a) 500 °C and (b) 1000 °C annealing. Log-Log plot of the integrated PL intensity as a function of excitation power density for each emission peak obtained after (c) 500 °C and (d) 1000 °C annealing.}
      \label{figureS_PD_PL}
    \end{figure}

Without any objective lens to focus the laser onto the sample, the excitation power density values were calculated for each of the ND filters used and are listed in Table \ref{table:unfocused laser power}. To estimate the unfocused laser spot size (within the interlocked enclosure) a 1 mm-spaced printed calibration grid was placed in its path. The laser spot illumination of this grid was then imaged using a smartphone camera yielding an estimated unfocused beam diameter of approximately 2 mm. Subsequently, the same PL spectrum measurements were performed with the CCS200 spectrometer and the corresponding data processing was repeated, with results shown in Figure \ref{figureS_PD_PL}. Note that any UVB emission, if present, is below the detection limit. For this reason, the UVB peak was excluded from the Gaussian fitting here and the subsequent rate equation modelling under the unfocused laser configuration.

    \begin{table}[!ht]
        \centering
        \caption{Excitation power density values of the unfocused laser for each ND filter used.}
        \begin{tabular}{ccc}
        \hline
            Power received (mW) & Nominal average power density ($\text{mW/cm}^2$) & Power percentage (\%)\\ \hline
            6.98 & 222.18 & 100 \\ 
            3.66 & 116.50 & 52.4 \\ 
            1.62 & 51.57 & 23.2  \\ 
            0.67 & 21.33 & 9.6 \\ 
            0.46 & 14.64 & 6.6  \\ 
            0.29 & 9.23& 4.2  \\ 
            66.5 $\times$ 10$^{-3}$ & 2.12 & 0.93  \\ 
            5.40 $\times$ 10$^{-3}$ & 0.17 & 0.08  \\ 
            0.62 $\times$ 10$^{-3}$ & 0.02 & 0.009  \\ \hline
        \end{tabular}
        \label{table:unfocused laser power}
    \end{table}

\section{Rate equation modelling under the unfocused laser configuration}
\label{SectionS6}

\begin{figure}[htbp]
      \centering
      \includegraphics[width=1\textwidth]{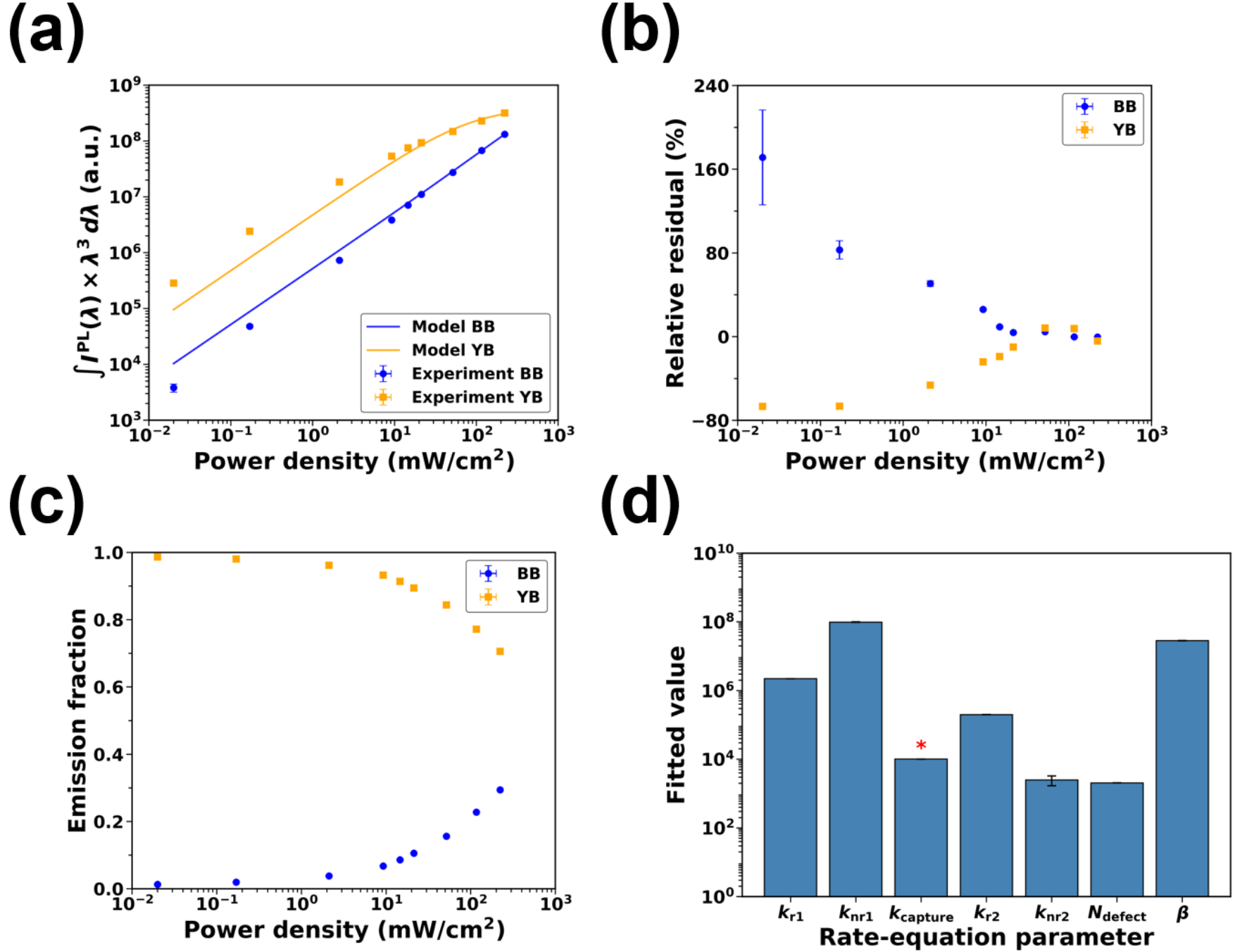}
      \caption{Summary of rate-equation modelling for unfocused laser excitation PL obtained after 500 °C annealing. (a) Comparison between the emission channels' experimental PL integrated intensities (data points) and the best fit obtained using the rate-equation model (lines) as a function of excitation power density. (b) Relative residuals between the emission channels' experimental and modelled PL integrated intensities. (c) Relative channel contribution of the experimental PL integrated intensities. (d) Mean fitting parameter values and standard deviation obtained from repeated differential evolution runs. An `*' identifies a parameter whose best-fit value lies within 5\% of a search-space boundary.}
      \label{figureS_REM_500}
    \end{figure}

\begin{figure}[htbp]
      \centering
      \includegraphics[width=1\textwidth]{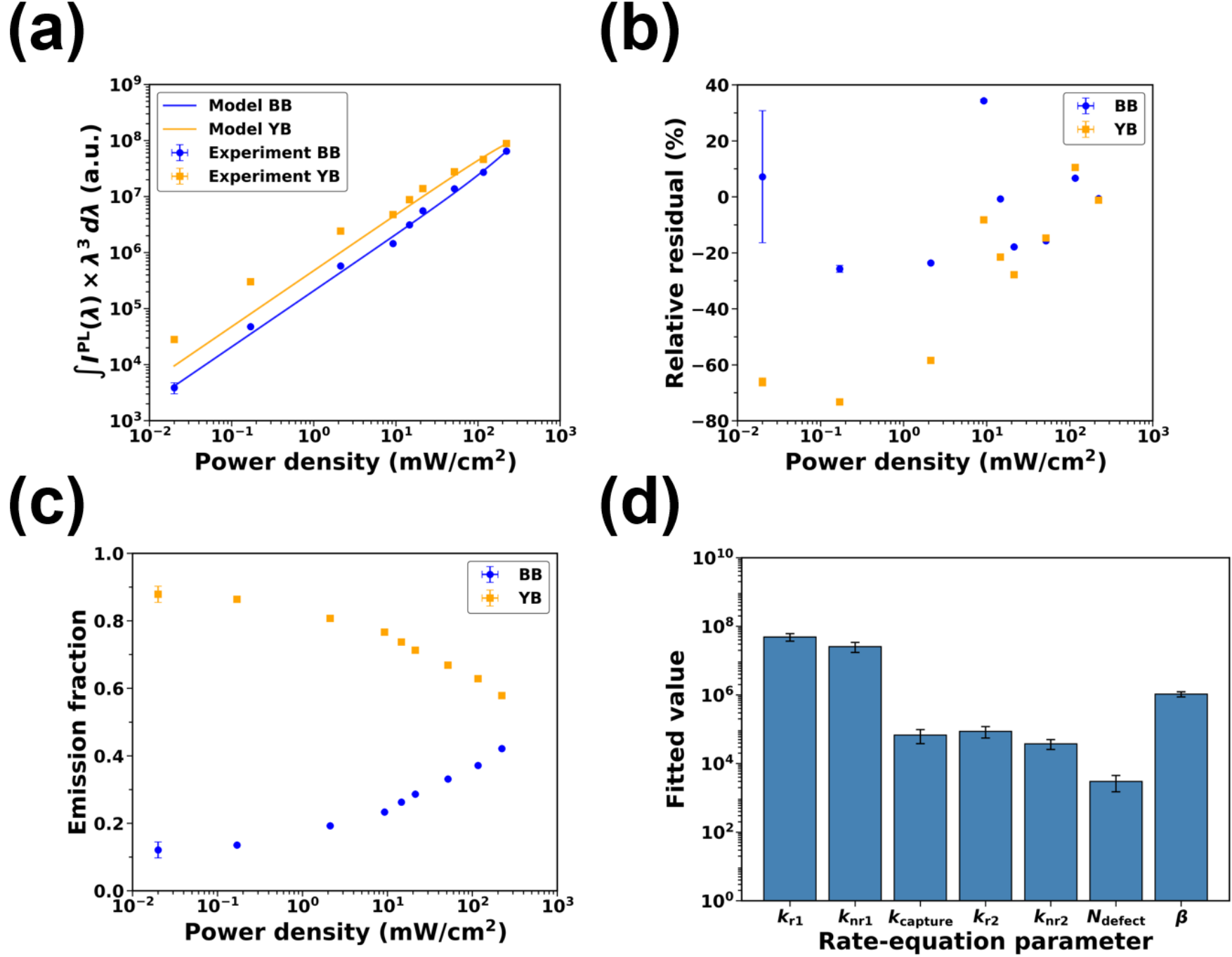}
      \caption{Summary of rate-equation modelling for unfocused laser excitation PL obtained after 1000 °C annealing. (a) Comparison between the emission channels’ experimental PL integrated intensities (data points) and the best fit obtained using the rate-equation model (lines) as a function of excitation power density. (b) Relative residuals between the emission channels’ experimental and modelled PL integrated intensities. (c) Relative channel contribution of the experimental PL integrated intensities. (d) Mean fitting parameter values and standard deviation obtained from repeated differential evolution runs.}
      \label{figureS_REM_1000}
    \end{figure}

The rate equation modelling was carried out for the excitation power density-dependent PL spectra under the unfocused laser configuration (without fitting for UVB emission), and the results are shown in Figures \ref{figureS_REM_500} and \ref{figureS_REM_1000}.

In a nonlinear fitting model, the coefficient of determination ($R^2$) is defined by Equation \ref{equ:coefficient of determination}, where $y_i^{\mathrm{Exp}}$, $y_i^{\mathrm{Model}}$, and $\overline{y}^{\mathrm{Exp}}$ respectively represent experimental quantities, modelled quantities, and the mean value of experimental quantities. $R^2$ describes the proportion of the total variance in the experimental quantities that can be explained by the modelled quantities. It is often used to estimate the goodness of fit, where a value of 1 suggests a perfect fit, while a value of 0 means the model is no better than a horizontal line passing through the mean value of experimental quantities. Table \ref{table:coefficient of determination} shows the distribution of $R^2$ values across all the fitting scenarios.
    
    \begin{equation}
        R^2 = 1 - \frac{\sum\limits_{i=1}^{n}\left(y_i^{\mathrm{Exp}} - y_i^{\mathrm{Model}}\right)^2}{\sum\limits_{i=1}^{n}\left(y_i^{\mathrm{Exp}} - \overline{y}^{\mathrm{Exp}}\right)^2}. \label{equ:coefficient of determination} 
    \end{equation}

    \begin{table}[!ht]
        \centering
        \caption{The coefficient of determination ($R^2$) values across all the fitting scenarios.}
        \begin{tabular}{ccc}
        \hline
            Conditions & $R^2$ for (UVB+BB) channel & $R^2$ for YB channel\\ \hline
            Post-annealed@500 °C, focused & 0.9955 & 0.9951 \\ 
            Post-annealed@1000 °C, focused & 0.9933 & 0.9987 \\ 
            Post-annealed@500 °C, unfocused & 0.9998 & 0.9879  \\ 
            Post-annealed@1000 °C, unfocused & 0.9974 & 0.9911  \\ \hline
        \end{tabular}
        \label{table:coefficient of determination}
    \end{table}

\section{AFM 2D topography and height profile of the ToF-SIMS sites}
\label{SectionS7}

\begin{figure}[htbp]
      \centering
      \includegraphics[width=0.8\textwidth]{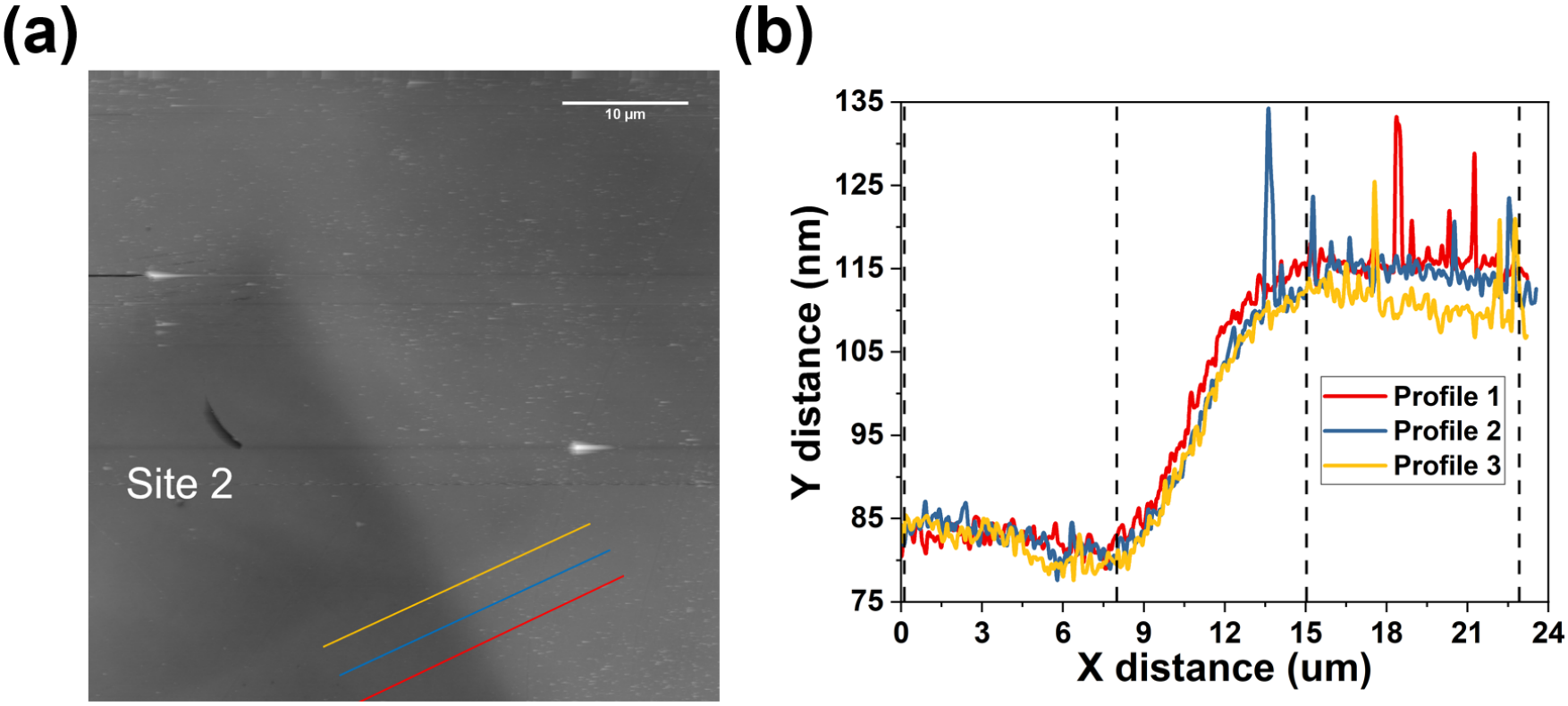}
      \caption{(a) The AFM topography of the site 2 ToF-SIMS crater boundary with three cross-sectional lines. (b) Corresponding depth profiles colour-coded to match the lines in (a).}
      \label{figureS_AlN_rate}
    \end{figure}

\begin{figure}[htbp]
      \centering
      \includegraphics[width=0.8\textwidth]{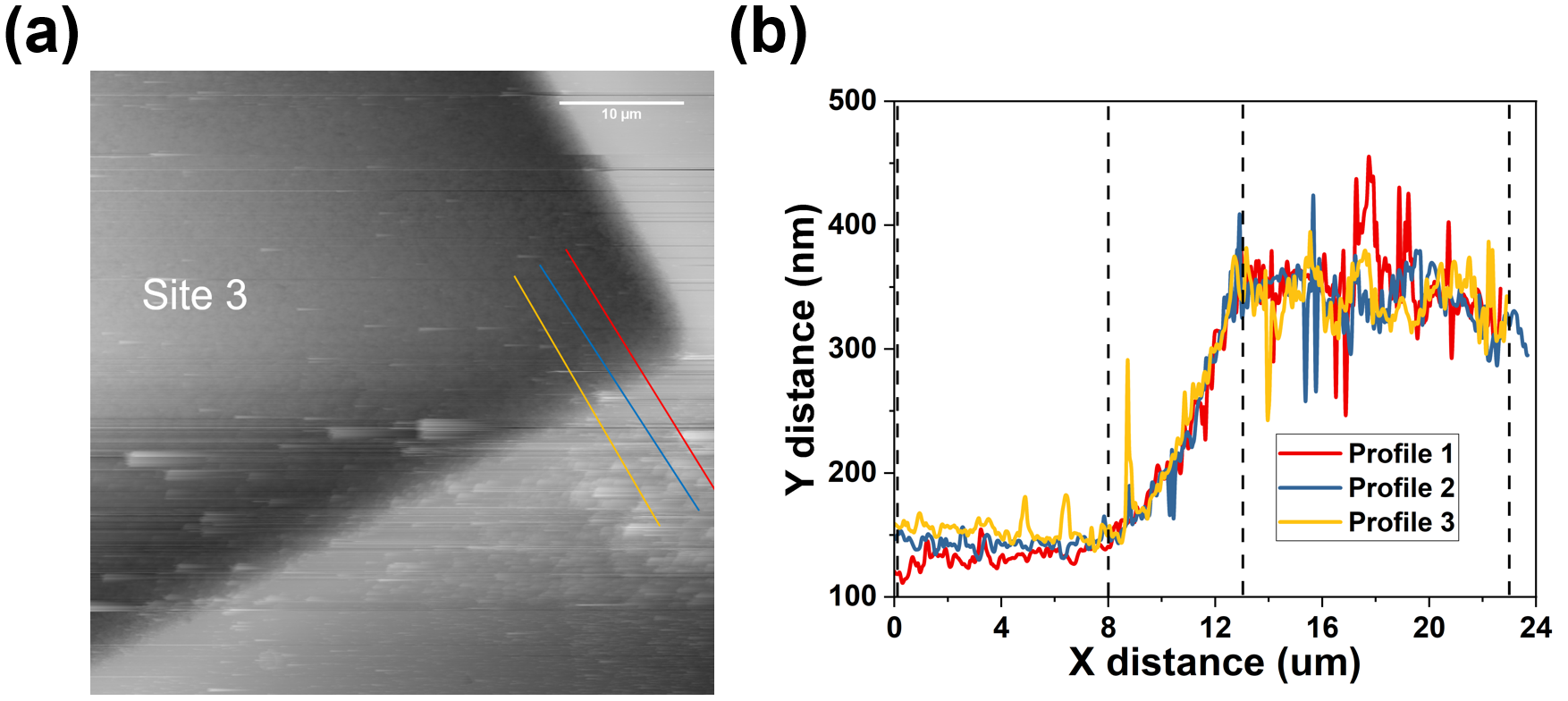}
      \caption{(a) The AFM topography of the site 3 ToF-SIMS crater boundary with three cross-sectional lines. (b) Corresponding depth profiles colour-coded to match the lines in (a).}
      \label{figureS_GaN_rate}
    \end{figure}

Following the same methodology mentioned previously, the step heights for the ToF-SIMS crater boundaries were determined to be 31.62 ±1.22 nm for site 2 (Figure \ref{figureS_AlN_rate}) and 201.43 ±10.22 nm for site 3 (Figure \ref{figureS_GaN_rate}), respectively.

%%% Uncomment this section and comment out the \bibliography{references} line above to use inline references.
% \begin{thebibliography}{1}

% 	\bibitem{kour2014real}
% 	George Kour and Raid Saabne.
% 	\newblock Real-time segmentation of on-line handwritten arabic script.
% 	\newblock In {\em Frontiers in Handwriting Recognition (ICFHR), 2014 14th
% 			International Conference on}, pages 417--422. IEEE, 2014.

% 	\bibitem{kour2014fast}
% 	George Kour and Raid Saabne.
% 	\newblock Fast classification of handwritten on-line arabic characters.
% 	\newblock In {\em Soft Computing and Pattern Recognition (SoCPaR), 2014 6th
% 			International Conference of}, pages 312--318. IEEE, 2014.

% 	\bibitem{hadash2018estimate}
% 	Guy Hadash, Einat Kermany, Boaz Carmeli, Ofer Lavi, George Kour, and Alon
% 	Jacovi.
% 	\newblock Estimate and replace: A novel approach to integrating deep neural
% 	networks with existing applications.
% 	\newblock {\em arXiv preprint arXiv:1804.09028}, 2018.

% \end{thebibliography}

\end{document}